\documentclass[manuscript,screen]{acmart}

\AtBeginDocument{%
  }

\usepackage[hang,flushmargin,para]{footmisc}

\usepackage{wrapfig}
\usepackage{tcolorbox}
\usepackage{tikz}
\usepackage{newfloat}
\usepackage{bm}
\usepackage{multirow}
\usepackage{booktabs}
\usepackage{svg} 
\usepackage{tabularx}
\usepackage{mathrsfs}
\usepackage{amsthm}
\usepackage{amsmath}
\usepackage{array}
\usepackage{xcolor, colortbl}
\usepackage{makecell}

\newcounter{papernumber}
\newcommand{\response}[1]{\noindent\textbf{Response:} #1} %
\newcommand{\revision}[1]{\noindent\textbf{Revision in the paper:} #1} %
\definecolor{softblue}{RGB}{0, 166, 100}
\definecolor{softyellow}{RGB}{124, 124, 186}
\definecolor{softcyan}{RGB}{63, 160, 192}

\begin{document}
\title{Graph-based Approaches and Functionalities in Retrieval-Augmented Generation: A Comprehensive Survey}

\author{Zulun Zhu}
\affiliation{%
  \institution{Nanyang Technological University}
  \country{Singapore}}
\email{ZULUN001@ntu.edu.sg}

\author{Tiancheng Huang}
\affiliation{%
  \institution{Nanyang Technological University}
  \country{Singapore}}
\email{tiancheng.huang@ntu.edu.sg}

\author{Kai Wang}
\affiliation{%
  \institution{Nanyang Technological University}
  \country{Singapore}}
\email{kai_wang@ntu.edu.sg}

\author{Junda Ye}
\authornote{This work was completed while Junda Ye was visiting Nanyang Technological University.}
\affiliation{%
  \institution{Beijing University of Posts and Telecommunications}
  \country{China}}
\email{jundaye@bupt.edu.cn}

\author{Xinghe Chen}
\affiliation{%
  \institution{Nanyang Technological University}
  \country{Singapore}}
\email{xinghe001@e.ntu.edu.sg}

\author{Siqiang Luo}
\authornote{Siqiang Luo is the corresponding author.}
\affiliation{%
  \institution{Nanyang Technological University}
  \country{Singapore}}
\email{siqiang.luo@ntu.edu.sg}

\renewcommand{\shortauthors}{Z. Zhu et al.}

\begin{abstract}
Large language models (LLMs) struggle with the factual error during inference due to the lack of sufficient training data and the most updated knowledge, leading to the {\it hallucination} problem. Retrieval-Augmented Generation (RAG) has gained attention as a promising solution to address the limitation of LLMs, by retrieving relevant information from external source to generate more accurate answers to the questions. Given the pervasive presence of structured knowledge in the external source, considerable strides in RAG have been made to employ the techniques related to graphs and achieve more complex reasoning based on the topological information between knowledge entities.  {Although recent surveys summarize RAG pipelines, they largely restrict the role of graphs to knowledge-graph traversal, leaving the broader impacts of graph structure on retrieval, prompting, and pipeline control underexplored.}
This survey offers a novel perspective on the functionality of graphs within RAG and their impact on enhancing performance across a wide range of graph-structured data. It provides a detailed breakdown of the roles that graphs play in RAG, covering database construction, algorithms, pipelines, and tasks. Finally, it identifies current challenges and outline future research directions, aiming to inspire further developments in this field.
Our graph-centered analysis highlights the commonalities and differences in existing methods, setting the stage for future researchers in areas such as graph learning, database systems, and natural language processing.

\end{abstract}

\ccsdesc[500]{Information systems~Information retrieval}
\ccsdesc[500]{Computing methodologies~Natural language processing}


\keywords{Graph, Large language model, Retrieval-augmented generation.}



\maketitle

\section{Introduction}
Large Language Models (LLMs) have demonstrated high effectiveness in a wide spectrum of human-centric tasks, including text summarization \cite{ding2024evaluation}, question answering \cite{ref:ewek-qa}, machine translation \cite{feng2024improving}, and code generation \cite{liu2024your}. Early models like BERT \cite{ref:bert} and RoBERTa \cite{ref:roberta} introduced advanced techniques for leveraging contextual embeddings, significantly enhancing performance of different natural language understanding tasks. Inspired by these fundamental works, models such as GPT-4~\cite{ref:gpt4}, LLaMA 2~\cite{ref:llama}, and more recently Deepseek-R1~\cite{ref:deepseek}, have set new benchmarks by utilizing transformer decoder-only architectures and scaling up model sizes. These LLMs excel at generating human-like responses, adapting to a wide range of domains, and supporting complex reasoning. All of these advancements have led to a growing consensus that LLMs represent a key pathway to artificial general intelligence (AGI)~\cite{bubeck2023sparks}.

Despite these successes, LLMs face two major challenges: they are prone to factual errors and hallucinations, and they frequently struggle with real-world knowledge that inherently represents structured information~\cite{lavrinovics2025knowledge, agrawal2023can, zhu2024kg}. The first challenge arises because the retraining of LLMs to incorporate new or domain-specific knowledge is computationally expensive \cite{ref:talk}, leaving models susceptible to producing logically coherent but factually incorrect inferences, which is also known as the hallucination problem \cite{Shuster2021retrieval}. For the second challenge, knowledge graphs like DBpedia \cite{dbpedia}, Wikidata \cite{ref:wikidata}, and Freebase \cite{ref:freebase} are primary examples of how entities and relationships are organized in a structured format, yet existing LLM-based approaches often fail to fully exploit these relational patterns, especially for complex or multi-hop reasoning tasks.

These two challenges call for retrieval mechanisms that enable LLMs to dynamically access \textbf{external} and \textbf{structured} knowledge during inference. Retrieval-Augmented Generation (RAG) helps solve the first challenges by retrieving relevant external information, thereby reducing hallucinations and enhancing factual accuracy \cite{ref:ragsurvey5, ref:ragsurvey3}. With respect to the second challenge, the integration of graph data management with RAG further enhances these capabilities by effectively organizing knowledge into organized forms such as knowledge graphs \cite{ref:GPT4Graph}. This integration captures complex relational patterns and supports multi-hop reasoning tasks such as node classification \cite{ref:Graph-LLM, ref:graphtranslator}, link prediction \cite{ref:wn, ref:kg-r3}, and subgraph extraction \cite{ref:hipporag, ref:rasr, ref:dalk}. Consequently, graph-enhanced RAG provides a powerful approach to improving LLM's factual reliability and their ability to execute complex, structured reasoning tasks.

 {The application of graph techniques in RAG has rapidly expanded, leading several recent surveys on LLMs to discuss graph-based methods, such as Han's survey, Peng's survey in 2024 \cite{peng2024graph} and Zhang et al.'s more recent work in 2025 \cite{zhang2025survey}. }
 {While prior surveys provide a useful synthesis of RAG architectures and applications with graphs, they are largely pipeline-centric and often conflate graph-based RAG with knowledge-graph traversal. Consequently, they devote limited attention to the underlying graph techniques, and provide mainly high-level rather than task-specific guidance for applying graph techniques in LLM pipelines. }
 {In contrast, this paper {fills this notable gap} by providing a meticulous and detailed decomposition of the RAG framework, broadening the definition of graph-based RAG and thoroughly examining the concrete graph techniques applied in representative works.} We explicitly discuss the rationale behind integrating these graph methods into RAG systems and highlight their specific advantages in enhancing answer accuracy and efficiency. Furthermore, based on insights drawn from over 200 recent RAG studies, we systematically categorize the use of graph techniques across the core components of RAG—including database construction, retrieval or prompting algorithms, processing pipelines, and task scenarios. Unlike previous surveys \cite{peng2024graph, ref:ragsurvey1, ref:ragsurvey2, ref:ragsurvey5}, this work offers an in-depth exploration from the viewpoint of graph data management with respect to database construction, query algorithms, prompt structure, and pipeline design,  presenting novel insights into how detailed graph methods can significantly drive advancements in RAG research. The contributions of this work are summarized as follows: 

$\bullet$ \textbf{Detailed and Case-by-Case Introduction of Graph Techniques in RAG:}
We meticulously introduce concrete graph techniques employed in RAG, analyzing them case by case, unlike the brief and superficial descriptions provided in existing surveys. This systematic and detailed review clarifies both the rationale for selecting these methods and their practical benefits in enhancing accuracy and efficiency.

$\bullet$ \textbf{Novel Perspective and Taxonomy:} We introduce a novel perspective on the roles of graphs in RAG, encompassing database construction, algorithms, pipelines, and tasks. Additionally, we propose a taxonomy to categorize existing RAG methods and provide insightful discussion of their their pros and cons.

$\bullet$ \textbf{Comprehensive Resource and Future Directions:} Based on an extensive review of over 200 studies, we provide a comprehensible resource to guide graph researchers in contributing to this evolving field. We introduce the representative works in detail to demonstrate the functionality of graphs to enhance the reasoning process of LLMs. 
This work lays the foundation for future advancements in areas such as graph learning, database systems, and natural language processing.

\section{Comparison with Existing Surveys}
\textbf{Surveys of LLMs.} 
A large number of surveys \cite{yang2023harnessing,huang2022towards,naveed2023comprehensive,zhao2023survey,minaee2024large,hadi2023survey,ref:llm_survey,ref:cot-survey} have explored LLMs from various perspectives over the years, reflecting their growing prominence in AI research. Early works \cite{yang2023harnessing,hadi2023survey} provide comprehensive overviews of foundational models like BERT \cite{ref:bert} and GPT-3 \cite{ref:gpt3}, detailing their architectures, design principles, and advancements compared to earlier models. These surveys emphasize the evolution of LLMs and their capacity to handle zero-shot and few-shot learning tasks, marking a shift from task-specific fine-tuning.
Other studies \cite{zhao2023survey,minaee2024large,shen2023large} focus on specific technical aspects, such as pretraining methods, fine-tuning strategies, and alignment techniques for generating reliable outputs. They also address challenges like mitigating biases, ensuring factual accuracy, and aligning outputs with user intent.


\noindent\textbf{Surveys of RAG.}
Building on the foundation of LLM surveys, an increasing number of works have concentrated on RAG, exploring its workflow and the integration of external knowledge to enhance generative accuracy. Some surveys \cite{ref:ragsurvey1, ref:ragsurvey2, ref:ragsurvey3, ref:ragsurvey5, ref:ragsurvey6,ref:ragsurvey7,llmir} delve into the architecture of RAG, detailing how retrieval mechanisms interact with generative models to create a seamless pipeline for producing context-aware responses. These studies often highlight the modularity of RAG workflows, emphasizing the flexibility of incorporating different retrieval strategies, such as vector-based retrieval and hybrid methods, to optimize performance across diverse tasks.
Other surveys \cite{tonmoy2024comprehensive,ref:hallucination, rawte2023survey,zhang2023siren} prioritize addressing the hallucination problem, analyzing how external information retrieval mitigates factual inaccuracies, and ensures the generated responses align closely with real-world knowledge. They discuss the effectiveness of external retrieval in outputs of grounding models and reducing the risk of plausible but incorrect information generation. Additionally, some works \cite{zhao2024retrieval,ref:ragsurvey4,zhou2024trustworthiness} explore the broader applications of RAG, demonstrating its utility in tasks like question answering, summarization, and domain-specific knowledge generation. However, given the importance of graphs as a fundamental structure in RAG, graph-related applications are not a central focus of these works.

\noindent\textbf{Surveys of LLMs on graphs.} 
Existing surveys of LLMs on graphs can broadly be categorized into two categories. The first category focuses on how LLMs can enhance prediction tasks on graphs by utilizing their ability to comprehend natural language \cite{ref:gfmsurvey1, ref:gfmsurvey2, ref:gfmsurvey3,ref:gfmsurvey4,ref:gfmsurvey7}. These surveys examine techniques such as (i) feature augmentation, where LLMs enhance node or edge attributes; (ii) feature alignment, which bridges textual and graph representations; and (iii) structural enhancement, leveraging LLMs to model intricate graph relationships and topological patterns. These works emphasize the potential of LLMs in boosting the performance of tasks like node classification, link prediction, and graph-based recommendation systems.
Surveys in another category \cite{ref:gfmsurvey5, ref:gfmsurvey6,ref:gfmsurvey8} investigate how LLMs enhance traditional symbolic knowledge bases through three key approaches: (i) deploying LLMs as knowledge graph builders and controllers; (ii) leveraging structured knowledge to improve LLM pretraining; and (iii) enabling LLM-augmented symbolic reasoning to refine logical inference. With the approaches above, these methods showcase the potential of LLMs to enrich and expand the capabilities of knowledge graphs in both construction and reasoning tasks. Another important related work \cite{peng2024graph} examines the general workflow of how RAG leverages external knowledge graphs to enhance LLM predictions.

 {\noindent\textbf{Comparison with existing surveys on graph-based RAG.} 
Existing surveys on graph-based RAG \cite{han2024retrieval, peng2024graph} provide a valuable synthesis of the RAG workflow and have helped stabilize terminology and component boundaries. At the same time, they are largely component-centric and tend to equate graph-based RAG with knowledge-graph traversal, which narrows the view of what graphs contribute to retrieval and reasoning. This perspective gives limited attention to analysis grounded in data-management practice and offers only modest, task-level guidance for graph researchers who wish to apply graph techniques in LLM-centric pipelines.}

 {Our survey addresses these gaps along three axes while building on the strengths of prior work. First, it adopts a broader definition of graph-based RAG, encompassing graph-powered databases beyond curated knowledge graphs and clarifying how structural signals inform retrieval and prompting. Second, it provides a deeper analysis from the perspective of graph data management, treating retrieval, prompting, and pipeline topology as design problems with explicit trade-offs in accuracy, latency, token budget, freshness, and provenance. Third, it serves as a comprehensible resource for graph researchers, offering task-wise guidance and decision criteria that indicate when graph structure materially improves outcomes and how to apply it in practice. Together, these contributions complement existing surveys by shifting the emphasis from pipeline components to the roles graphs play in RAG and by translating those roles into actionable methodology.
}





\section{Foundations of Graph-Enhanced RAG}

\begin{figure*}
    \centering
\includegraphics[width=4in]{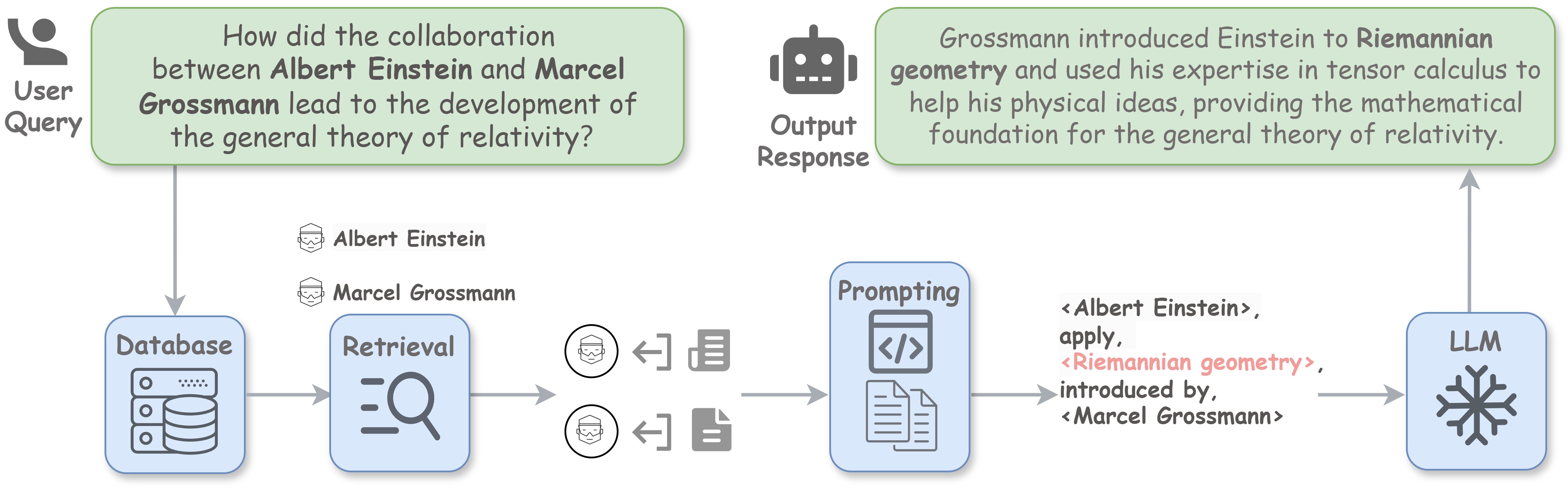}
\vspace{-2mm}
    \caption{Paradigm overview of RAG}
    \vspace{-5mm}
    \label{fig:rag}
\end{figure*}

\subsection{Overall Workflow of RAG}
The standard RAG pipeline combines external knowledge retrieval with the generation abilities of LLMs, creating a robust pipeline for producing contextually accurate responses.  
We present an example for demonstration in Figure \ref{fig:rag}. In this scenario, a user query explores the development of the \textit{general theory of relativity}, which involves background knowledge about scientists \textit{Albert Einstein} and \textit{Marcel Grossmann}. As this information may exceed the capacity of an LLM, RAG extends its scope by retrieving relevant knowledge from an external database. This process constructs a more informative prompt, thereby improving response accuracy. The complete workflow can be divided into the following four steps:

\textbf{(a) User query.}
The workflow begins with a user-provided query, typically in natural language, such as a question or request for specific information. In our running example, the user initiates a dialogue to explore the development of the \textit{general theory of relativity}. Additionally, the query includes a constraint specifying that the response should focus on the collaboration between \textit{Albert Einstein} and \textit{Marcel Grossmann}.

\textbf{(b) Retrieval module.} The query is passed to a retrieval system that interacts with an external knowledge database, such as a document corpus, knowledge graph, or vector store. Then, the retrieval system identifies and fetches the information most closely aligned with the query. The example query involves three key entities essential for answering the question: \textit{general theory of relativity}, \textit{Albert Einstein}, and \textit{Marcel Grossmann}. The RAG system retrieves relevant information associated with these entities. In the context of a knowledge graph, the retrieved knowledge can be represented as triples, such as <\textit{Marcel Grossmann}, introduced \textit{Riemannian geometry} to, \textit{Albert Einstein}>.

\textbf{(c) Prompting module.} The retrieved knowledge is transformed and integrated into the original query, ensuring that the model has access to up-to-date and domain-specific knowledge for generating accurate responses. Based on the key information retrieved to address the development of the \textit{general theory of relativity}, the RAG system integrates external knowledge and reformats it to align with the LLM's input style. For example, it may generate a statement such as: "\textit{Albert Einstein} applied \textit{Riemannian geometry} in developing the \textit{general theory of relativity}, a concept introduced to him by \textit{Marcel Grossmann} during their collaboration." This prompting strategy helps the LLM more effectively process and incorporate new information naturally into its responses.

\textbf{(d) Output response.} Finally, the LLM leverages the enriched context to generate a response that is both accurate and highly relevant to the user’s query. By incorporating external knowledge, the model not only captures the core rationale—such as the role of \textit{Riemannian geometry} in the development of the \textit{general theory of relativity}—but also structures the information in a more coherent and explanatory manner. This ensures that the response not only conveys factual accuracy but also presents the historical and conceptual development of the \textit{general theory of relativity} in a clearer, more logically connected way.

\subsection{RAG with Graph Data Management}
\begin{figure*}
    \centering
\includegraphics[width=0.7\linewidth]{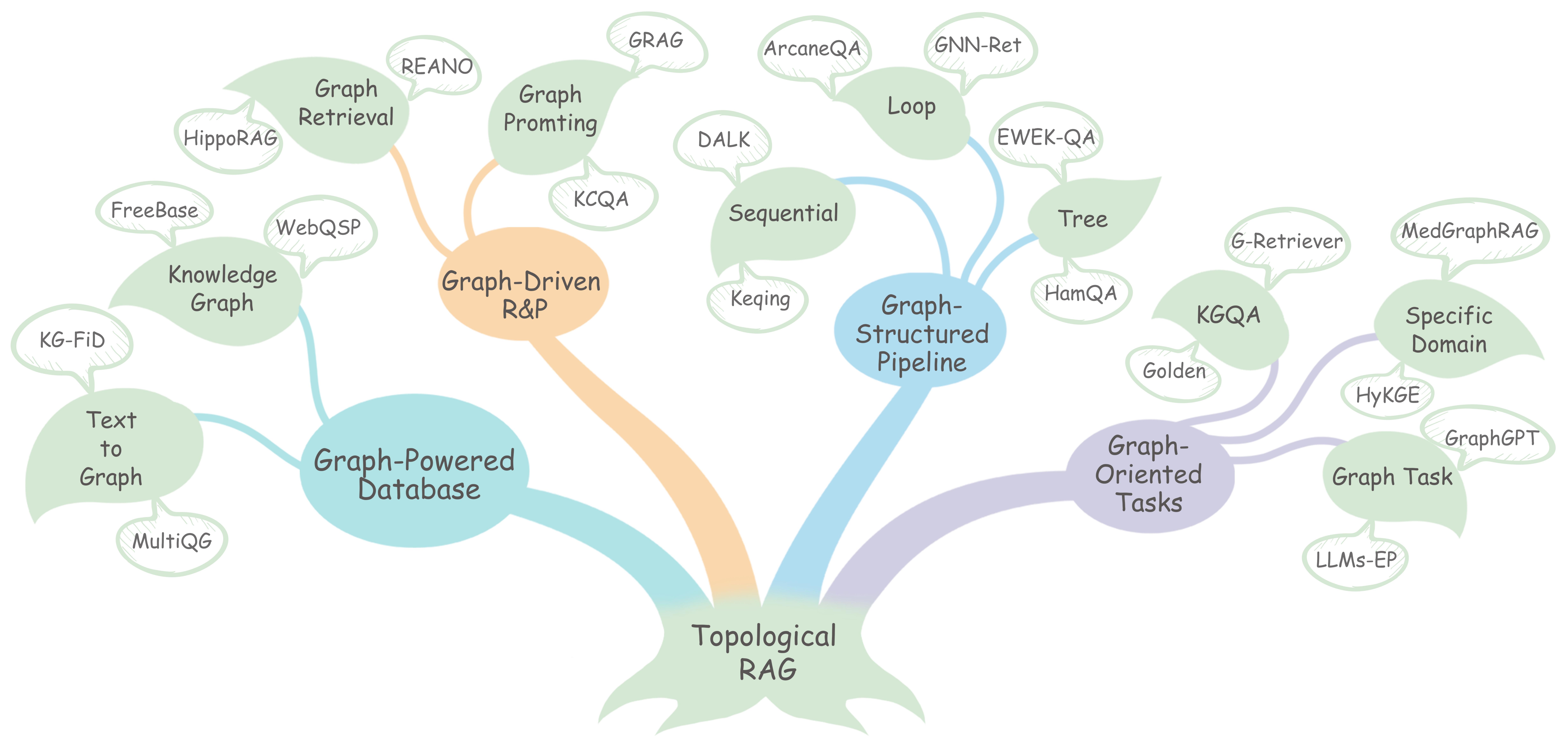}
\vspace{-2mm}
    \caption{RAG with graph data management}
    \vspace{-4mm}
    \label{fig:overall}
\end{figure*}

Graphs, with their inherent ability to model complex relationships and dependencies, serve as powerful tools for structuring external knowledge and enabling sophisticated reasoning in RAG workflows. 
Existing literature \cite{ref:gfmsurvey7,peng2024graph}  primarily focuses on how LLMs enhance graph performance or provide a brief introduction of the RAG process involving graphs. However, several critical questions remain unanswered: \textit{What benefits does the graph structure offer to RAG? Which graph patterns can enhance the effectiveness or efficiency of LLMs responses? What future directions should be pursued to develop graph algorithms that synergize with LLMs? } To address these compelling questions, we approach this survey explicitly from the perspective of graph data management, systematically investigating how graph techniques enhance each core stage of the RAG process, as illustrated in Figure \ref{fig:overall}. Specifically, we highlight how graph data management methods significantly (i) improve knowledge base construction; (ii) optimize retrieval and prompting algorithms through graph-based indexing, querying, and reasoning; (iii) streamline data processing pipelines via graph-structured workflows; and (iv) effectively support graph-oriented tasks. 

 \textbf{Graph-Powered Databases.} 
In Section \ref{sec:database}, We will explore the methods used to construct auxiliary databases for RAG. The discussion will be organized into two categories: existing knowledge graphs and the graph from the texts. For each category, we will analyze the representative graph techniques employed in existing literature and examine how they contribute to building effective auxiliary databases. This analysis will provide insights into how these databases are prepared to support subsequent retrieval tasks in RAG systems.


 \textbf{Graph-Driven R\&P (retrieval and prompting).} In Section \ref{sec:RP}, we will examine the graph algorithms used in RAG from two perspectives: graph retrieval and prompting (R\&P). For graph retrieval, we will categorize existing graph techniques into non-parameterized methods and learning-based methods, analyzing their effectiveness and efficiency in retrieving relevant information. For Graph Prompting, we will explore two distinct classes: Graph-Structure Prompting and Text Prompting, focusing on how they transform graph information into formats that can be utilized by LLMs. For each category, we present concrete examples to illustrate the specific techniques and their practical applications.


 \textbf{Graph-Structured Pipelines.} In Section \ref{sec:pipeline}, we categorize the pipelines in  RAG into three distinct types: Sequential Pipeline, Loop Pipeline, and Tree Pipeline. This classification is based on the way where each method structures the overall pipeline, reflecting the underlying graph-like relationships between different module. By capturing the topological characteristics of these pipelines, we aim to shed light on how their designs influence efficiency, scalability, and reasoning capabilities. Through case studies of representative projects, we will analyze how these different pipeline facilitate the integration of retrieval and generation, ultimately enhancing the performance of LLMs in diverse scenarios.

  \textbf{Graph-Oriented Tasks.} Finally, in Section \ref{sec:task}, we systematically categorize the applications of RAG into three key tasks: (i) knowledge graph question answering (KGQA) tasks, which involves responding to natural language queries using structured knowledge; (ii) graph-centric tasks (e.g., node classification, link prediction), leveraging reasoning property of LLMs to enhance graph reasoning; and (iii) domain-specific applications (e.g., biomedicine, finance), where graphs integrate domain knowledge to refine retrieval precision. For each category, we analyze how different RAG frameworks harness graph structures to optimize retrieval mechanisms, improve contextual relevance, and strengthen inferential accuracy.

\section{Graph-Powered Databases}\label{sec:database}
{Database construction serves as a foundational step in the RAG paradigm, as it organizes and stores external knowledge to facilitate effective information retrieval. Specifically, the database in graph-based RAG captures entities and relationships from plain textual knowledge into a structured format, enabling efficient access to local (immediate neighbors or direct relationships) and global (overall connectivity or multi-hop paths) topological information. Formally, let $\mathcal{E}$ denote the set of entities and $\mathcal{R}$ denote the set of relationships; a particular fact is represented as a triplet: $\mathbf{t} = (e_h, r, e_t) \in \mathcal{E}\times \mathcal{R}\times \mathcal{E}$, where $e_h$ and $e_t$ denote the head and tail entities, respectively, and $r$ denotes the relationship linking them. A graph database $\mathcal{G}$ thus consists of a collection of such triples: $\mathcal{G} = {(e_h, r, e_t)}.$ In practice, various types of graph databases have been utilized in the RAG literature (summarized in Table~\ref{table:dataset}), each offering distinct characteristics suited to different retrieval and reasoning tasks. In the following sections, we first introduce the existing graph databases and then discuss widely adopted techniques for generating such databases from textual data. }

\subsection{Existing Knowledge Graphs} 

\subsubsection{A Unified View}
{In this section, we focus on existing knowledge graphs, which play a crucial role in many RAG systems by serving as structured repositories of factual knowledge. To enable efficient querying and traversal of entities and their relationships, many existing methods \cite{ref:grag,ref:kaping,ref:unikgqa} directly retrieve knowledge relevant to the query from well-established knowledge graphs such as FreeBase \cite{ref:freebase}, T-REx \cite{ref:t-rex}, and WebQSP \cite{ref:webqsp}. These widely used databases offer comprehensive scopes for knowledge retrieval, encompassing a vast array of entities and relationships across diverse domains. For instance, T-REx \cite{ref:t-rex} provides extensive alignments between natural language expressions and knowledge base triples, facilitating the integration of structured data with textual information. By organizing nodes and edges along with their textual attributes, these structured repositories allow RAG systems to efficiently access additional topological information, supporting advanced operations such as entity retrieval, relationship discovery, and complex multi-hop queries. } 


\subsubsection{Existing Knowledge Graphs} 

\begin{table}[htb]
\centering
\small
\vspace{-5mm}
\caption{Summary of commonly used KGs and the methods constructing KG.}
\vspace{-2mm}
\begin{tabular}{m{1.5cm}<{\centering}|m{1.5 cm}<{\centering}|m{1.5 cm}<{\centering}|m{8.8 cm}<{\centering}} \toprule[1pt]
\textbf{Categories} & \textbf{Pros}& \textbf{Cons}&\textbf{References}\\  \midrule[0.5pt]
Existing KGs& \textit{Broad coverage, reliable quality}&\textit{General knowledge, less adaptable}&BioRED~\cite{ref:biored}, QALD-9-plus~\cite{ref:qald}, OpenbookQA~\cite{ref:openbookqa}, CREAK~\cite{ref:creak}, TriviaQA~\cite{ref:triviaqa}, HotpotQA~\cite{ref:hotpotqa}, Mintaka~\cite{ref:mintaka}, MedQA~\cite{ref:medqa},
TUDataset~\cite{ref:tudataset},
CWQ~\cite{ref:cwq}, Beyond I.I.D. ~\cite{ref:grailqa}, CommonsenseQA~\cite{ref:commonsenseqa}, SocialIQA~\cite{ref:socialIQA}, PIQA~\cite{ref:piqa}, RiddleSense~\cite{ref:riddle}, Freebase~\cite{ref:freebaseqa}, ATOMIC~\cite{ref:atomic}, FactKG~\cite{ref:factkg},
MultiHop-RAG~\cite{tang2024multihop}, T-REx~\cite{ref:t-rex}, DBpedia~\cite{ref:dbpedia}, Yago~\cite{ref:yago}\\ \midrule[0.5pt]
 Graphs generated from texts & \textit{Domain-specific, easily updated}& \textit{LM-dependent, computationally intensive}&GraphRAG~\cite{ref:graphrag}, GRBK \cite{ref:noname5}, ATLANTIC~\cite{ref:atlantic}, GNN-Ret~\cite{ref:gnn-ret}, HippoRAG~\cite{ref:hipporag}, DALK~\cite{ref:dalk}, KGP~\cite{ref:kgp}, OpenSCR~\cite{ref:opencsr}, MindMap~\cite{ref:mindmap}, FABULA~\cite{ref:fabula}, GER~\cite{ref:ger}, FoodGPT~\cite{ref:foodgpt}, ChatKBQA~\cite{ref:chatkbqa}, MultiQG~\cite{ref:noname8}, HSGE~\cite{ref:hsge}, ReTraCk~\cite{ref:retrack}, RNG-KBQA~\cite{ref:rng}, ArcaneQA~\cite{ref:arcaneqa}, HybridRAG~\cite{ref:hybridrag}, EWEK-QA~\cite{ref:ewek-qa}, KG-FiD~\cite{ref:kgfid}, REANO~\cite{ref:reano}, MedGraphRAG~\cite{ref:medgraphrag}, MINERVA~\cite{ref:minerva}\\
\bottomrule[1pt]
\end{tabular}
\vspace{-2mm}
\label{table:dataset}
\end{table}

{\textbf{Freebase} \cite{ref:freebase} is a large-scale, structured knowledge database designed to organize and store general human knowledge. It combines the efficiency and scalability of structured databases with the rich semantic representation found in sources like Wikipedia \cite{bridge2001wikipedia}, enabling flexible integration of diverse knowledge. By structuring information into tuples, Freebase facilitates efficient querying, entity linking, and knowledge retrieval across multiple domains.} Freebase features an HTTP-based API powered by the Metaweb Query Language (MQL) \cite{flanagan2007developing}, which facilitates intuitive object-oriented queries and schema evolution. Its complete normalization philosophy ensures unique global identifiers (GUIDs) for real-world entities, while its lightweight typing system supports multiple data.
In the context of RAG, Freebase serves as a powerful auxiliary database, enabling methods to retrieve structured knowledge for enhanced reasoning and answer generation. With its ability to handle vast datasets and diverse relationships, Freebase provides a robust foundation for integrating external knowledge into RAG systems, improving their capability to process complex queries and generate context-aware responses.

\textbf{T-REx} \cite{ref:t-rex} is a large-scale dataset designed to align natural language from Wikipedia abstracts with knowledge base (KB) triples from Wikidata. Addressing the alignment problem, T-REx connects unstructured text with structured knowledge representations by providing 11 million triples aligned with over 3 million Wikipedia abstracts, covering more than 600 unique predicates. Its customizable alignment pipeline incorporates techniques like predicate linking, co-reference resolution, and distant supervision to create high-quality alignments. Compared to existing datasets, T-REx provides a larger scale, broader predicate coverage, and high accuracy. This dataset is instrumental in advancing tasks like relation extraction, KB population, and question answering, and serves as a foundational resource for RAG systems to retrieve structured knowledge and improve alignment for complex reasoning.

\begin{figure*}
    \centering
\includegraphics[width=0.7\linewidth]{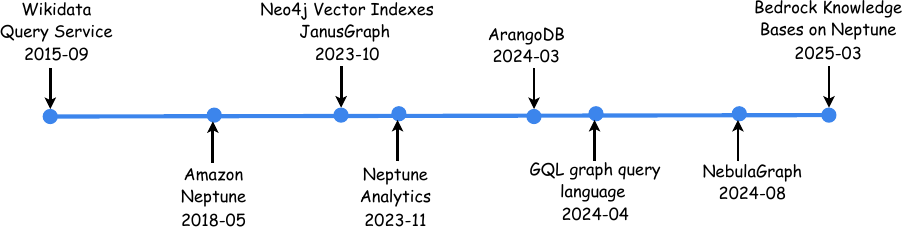}
\vspace{-2mm}
    \caption{Timeline of recent graph database}
    \vspace{-4mm}
    \label{fig:data_timeline}
\end{figure*}

   {
\subsubsection{Industrial Graph Database for LLM}
Recently, widely used graph databases have moved from early research backends to platforms that support LLM use cases with built-in embedding search, standard query languages, and managed GraphRAG pipelines. As shown in Figure \ref{fig:data_timeline}, we list these important databases on a timeline: Wikidata's public query service as a common open graph backend\footnote{{https://query.wikidata.org/}}; Amazon Neptune as a managed graph service, later adding Neptune Analytics for large-scale graph and embedding analytics\footnote{https://aws.amazon.com/cn/neptune/}; Neo4j with mature property-graph capabilities and native vector indexes for hybrid retrieval\footnote{https://neo4j.com/release/neo4j-4-0/}; JanusGraph as common open-source choices for large or real-time workloads\footnote{https://janusgraph.org/}; ArangoDB as a multi-model system that added vector indexes\footnote{https://www.arangodb.com/docs/3.12/release-notes/}; vendors beginning to productize GraphRAG (for example, NebulaGraph; Bedrock Knowledge Bases on Neptune)\footnote{https://www.nebula-graph.io/posts/announcing\_nebulagraph\_RAG}; and the GQL graph query language becoming an ISO standard that aligns vendor ecosystems\footnote{https://www.iso.org/standard/76120.html}.
}

   {
\subsubsection{Trend of Graph Database Construction}
In the LLM era, graph databases have evolved along four directions that directly address freshness, cost of construction, and LLM integration. First, \emph{standardization} has begun to reduce vendor lock-in and smooth model portability. For example, the publication of graph query language (e.g., ISO/IEC 39075:2024 \footnote{https://opencypher.org/, https://www.iso.org/standard/76120.html}, which formalizes a property-graph data and query language and enables more consistent schemas, queries, and tooling across systems.
Second, \emph{hybrid graph–vector indexing} has become more popular so that semantic similarity can be combined with graph traversal. For instance, Neo4j made vector indexes generally available and documents RAG workflows on top of them \footnote{https://neo4j.com/docs/cypher-manual/5/indexes/semantic-indexes/vector-indexes/}, while ArangoDB introduced native vector indexes and vector search functions of its declarative language \footnote{https://docs.arangodb.com/stable/}. Third, to lower \emph{update latency and construction overhead}, change-data-capture (CDC) and streaming ingestion became first-class, such as Neo4j’s CDC feature and Kafka connector strategies that propagate graph updates as events, reducing the gap between source changes and graph availability for RAG/agent pipelines \footnote{https://neo4j.com/blog/developer/change-data-capture-cdc-ga/}. Finally, the database construction increasingly blends \emph{general encyclopedic KGs} with \emph{domain KGs} for higher-quality grounding in graph-based RAG systems, a trend visible in community benchmarks such as GR-Bench \cite{ref:graphcot} and GraphQA \cite{ref:g-retriever}. In summary, these updates reduce data-to-serving latency and integration cost, making graph databases easier to operate at scale while preserving the multi-hop structure and provenance required by LLM applications.
}

\subsection{Graphs Generated from Texts}
\subsubsection{A Unified View}
\begin{figure*}
    \centering
\includegraphics[width=1\linewidth]{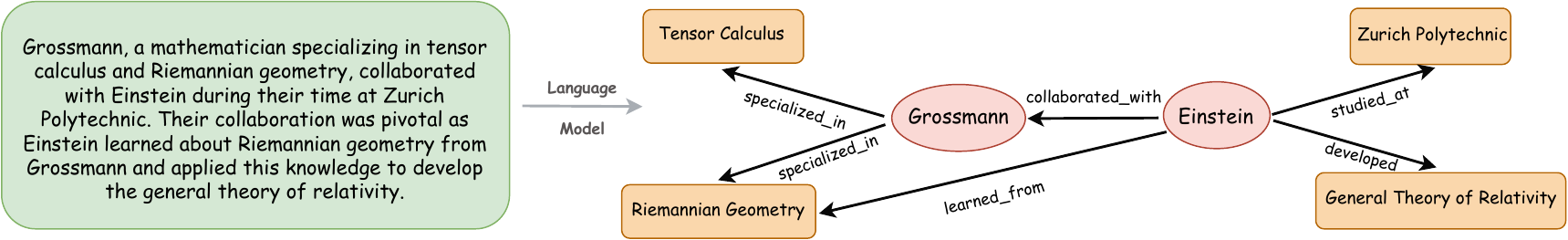}
\vspace{-2mm}
    \caption{From texts to knowledge graph}
    \vspace{-4mm}
    \label{fig:graph_to_text}
\end{figure*}

   {In addition to reusing existing knowledge graphs, we consider a broader class of graph-powered databases constructed from text. } A common route is transforming plain text into a knowledge graph through artificial means, which is known as open information extraction (OpenIE) \cite{angeli2015leveraging,etzioni2008open,zhou2022survey}. {This process extracts key information—such as entities, relationships, and contextual meanings—from textual documents. An instruction-tuned language model then identifies these entities and their relationships to generate structured triples.} As illustrated in Figure \ref{fig:graph_to_text}, given a paragraph describing the collaboration between \textit{Albert Einstein} and \textit{Marcel Grossmann}, the instruction-tuned LM captures the relationships among entities and converts them into structured triples. In longer contexts, multiple relationships emerge, linking entities to various neighbors and ultimately forming a complex graph. 

   {Beyond vanilla knowledge graphs, we also include multiple forms graphs distilled from text: (1) document-level structures such as entity–sentence and sentence–sentence links (HippoRAG \cite{ref:hipporag}, ATLANTIC \cite{ref:atlantic}, GNN-Ret \cite{ref:gnn-ret}, OpenSCR~\cite{ref:opencsr}, FiD~\cite{ref:kgfid}, MedGraphRAG~\cite{ref:medgraphrag}); (2) hierarchical graphs with each node as a subgraph (DALK \cite{ref:dalk}, GER~\cite{ref:ger}); (3) rule-based graph where each node aggregates multiple entities or facts consistent with a defined rule schema (FABULA~\cite{ref:fabula}, StarRAG \cite{StarRAG}, ReTraCk~\cite{ref:retrack}). These constructions produce property graphs with additional information and traceable evidence, enabling unique feature extractions from the knowledge base.
}
 


\subsubsection{Graph Generation Methods Based on Texts}
{\textbf{KG-FiD} \cite{ref:kgfid} is a graph-based framework that constructs knowledge graphs directly from textual data to facilitate more effective information retrieval. Specifically, it begins by extracting text segments from the dataset and linking them to their associated entities. }Using the BERT model for passage encoding, KG-FiD computes similarity scores and reranks the passages within entities to enhance retrieval accuracy. Additionally, a Graph Attention Network (GAT) is  utilized to propagate information across the graph, leveraging relationships between entities to retrieve relevant node information. This approach ensures that the graph representation is both contextually rich and informative, leading to improved retrieval outcomes.

\textbf{MultiQG} \cite{ref:noname8} proposes a method to answer multi-hop complex queries over knowledge bases by generating query graphs with constraints. The approach integrates constraints early in the process to guide the construction of query graphs, effectively reducing the search space. Candidate query graphs are ranked based on their embedding similarity with the question using a neural network, ensuring accurate alignment. By combining beam search and semantic matching, the method attains state-of-the-art performance on standard benchmark datasets like ComplexWebQuestions, demonstrating significant improvements in precision and F1 scores. This framework addresses the limitations of traditional RAG systems, which struggle to handle constraints effectively, improving accuracy in complex question.

\subsection{Comparison between Existing and Generated Knowledge Graphs }

Both existing knowledge graphs and text-derived graphs serve as essential components of graph-powered databases, each with distinct advantages and limitations. Existing knowledge graphs offer structured, high-precision information with efficient querying, making them well-suited for tasks requiring fast and reliable entity retrieval. However, they are inherently static and require manual updates to incorporate new knowledge. In contrast, graphs constructed from text provide greater adaptability by dynamically extracting relationships from unstructured data. While this approach enhances knowledge coverage and supports evolving information, it introduces potential workload due to reliance on language models and probabilistic extraction methods. Moving forward, hybrid approaches that integrate raw graphs with text-derived structures could offer a balanced solution, leveraging both efficiency and adaptability. Additionally, improving incremental learning for knowledge graph updates, enhancing entity-linking accuracy, and developing scalable retrieval mechanisms will be crucial for optimizing graph-powered databases.




\section{Graph-Driven R\&P}\label{sec:RP}
In the context of RAG, graphs play a crucial role in both retrieval and prompting (R\&P) for LLMs-- enabling the retrieval of structured knowledge and facilitating the prompting of LLMs. 
Specifically, \textit{graph retrieval} refers to the algorithms designed to extract additional knowledge from the graph, which serves to enrich the reasoning capabilities of the LLMs. This process involves traversing the graph to find relevant nodes, edges, or subgraphs that provide valuable context or information necessary to enhance the accuracy and depth of the model's responses. On the other hand, \textit{graph prompting} focuses on transforming the retrieved graph structure into a textual format that is easily understood by the LLMs. This transformation ensures that the complex relationships and connections within the graph are effectively communicated in natural language, enabling the LLMs to leverage the structured knowledge in its generative process. In the subsequent sections, we explore a range of graph-based retrieval and prompting techniques, highlighting representative approaches. The broader set of approaches, along with our classification, is presented in Table \ref{table:algorithm} and Table \ref{table:algorithm_prompt}.

\begin{table}[htb]
\centering
\footnotesize
\caption{Summary of retrieval algorithms.}
\vspace{-2mm}
\begin{tabular}{m{1.8 cm}<{\centering}|m{1.5 cm}<{\centering}|m{2 cm}<{\centering}|m{2 cm}<{\centering}|m{6cm}<{\centering}} 
\toprule[1pt]
\multicolumn{2}{c|}{\textbf{Retrieval Algorithms}} & \textbf{Pros}& \textbf{Cons}& \textbf{References}\\  \midrule[0.5pt]
&Deterministic Algorithms&\textit{Precise, reliable} & \textit{Intensive computation}&KG-GPT~\cite{ref:kg-gpt}, GLBK \cite{ref:noname5}, GRAG~\cite{ref:grag}, HyKGE~\cite{ref:hykge}, Knowledge Solver~\cite{ref:ksl}, KnowledGPT~\cite{ref:knowledgpt}, GGE~\cite{ref:gge}, Engine~\cite{ref:engine}, GraphBridge~\cite{ref:graphbridge}, MMGCN~\cite{ref:mmgcn}, NLGraph~\cite{ref:nlgraph}, GraphEval2000~\cite{ref:GraphEval2000} \\ \cmidrule(lr){2-5}
{\multirow[c]{3}{*}{\makecell[c]{Non-parameterized \\ Algorithms}}}&Probabilistic Algorithms & \textit{Scalable, strong adaptability}& \textit{Less precise, less interpretable}&HippoRAG~\cite{ref:hipporag}, MindMap~\cite{ref:mindmap}, OreoLM~\cite{ref:oreolm}, MultiQG~\cite{ref:noname8}, MINERVA~\cite{ref:minerva}, MuseGraph~\cite{ref:musegraph}, Walklm~\cite{ref:walklm}\\ \cmidrule(lr){2-5}
&Heuristic-Based Algorithms & \textit{Efficient, scalable}&\textit{Uncertain results depend on heuristics}& RA-SIM~\cite{ref:ra-sim}, KG-Rank~\cite{ref:kg-rank}, SKP~\cite{ref:skp}, NuTrea~\cite{ref:nutrea}, Zeshel~\cite{ref:zeshel}, Conll~\cite{ref:conll}, RNG-KBQA~\cite{ref:rng}, ArcaneQA~\cite{ref:arcaneqa}, HybridRAG~\cite{ref:hybridrag},MedGraphRAG~\cite{ref:medgraphrag}, TOG2~\cite{ref:tog2}, DepsRAG~\cite{ref:depsrag}, KELP~\cite{ref:kelp}, KGQA~\cite{ref:kgqa}, Graph-LLM~\cite{ref:Graph-LLM}, GLBench~\cite{ref:GLBench}
\\\midrule[1pt]
{\multirow[b]{2}{*}{\makecell[c]{Learning-based \\ Algorithms}}}&Convolutional-based Algorithms & \textit{Efficient local structure modeling} & \textit{Limited global information}& GNN-RAG~\cite{ref:gnn-rag}, MHKG~\cite{ref:feng}, RA-SIM~\cite{ref:ra-sim}, GenKGQA~\cite{ref:gentkgqa}, GRAG~\cite{ref:grag}, GNN-Ret~\cite{ref:gnn-ret}, ETD~\cite{ref:etd}, NuTrea~\cite{ref:nutrea}, KAM-CoT~\cite{ref:kamcot},  ConvE~\cite{ref:wn}, REANO~\cite{ref:reano}, SURGE~\cite{ref:SURGE}, GraphGPT~\cite{ref:GraphGPT}
\\ \cmidrule(lr){2-5}
&Attention-Based Algorithms& \textit{Global structure modeling}&\textit{High computational cost}&ATLANTIC~\cite{ref:atlantic}, OpenSCR~\cite{ref:opencsr}, GGE~\cite{ref:gge}, GER~\cite{ref:ger}, HSGE~\cite{ref:hsge}, HamQA~\cite{ref:hamqa},  KG-FiD~\cite{ref:kgfid}, Fact~\cite{ref:fact}
\\ 
\bottomrule[1pt]
\end{tabular}
\vspace{-4mm}
\label{table:algorithm}
\end{table}

\subsection{Retrieval with Non-parameterized Algorithms}

\subsubsection{A Unified View}
\begin{figure*}
    \centering
\includegraphics[width=0.9\linewidth]{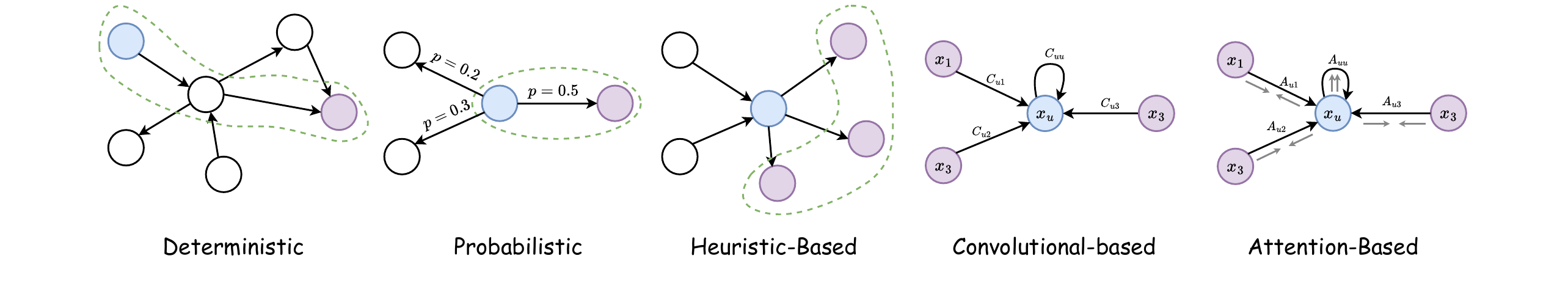}
\vspace{-2mm}
    \caption{Non-parameterized Algorithms and learning-based for Graph Retrieval}
    \vspace{-4mm}
    \label{fig:non learn alg}
\end{figure*}

The non-parameterized methods in the retrieval process of RAG focus on extracting relevant knowledge from the graph using predefined rules, without involving any trainable parameters. Specifically, given the query $q$ and the knowledge graph $\mathcal{G}$, the retrieval response $R_q$ from the graph can be formulated as:
$R_q = f(\mathcal{G}, e(q)),$
where $e(\cdot)$ is the entity or relationship extraction function and $f(\cdot)$ is a deterministic function that follows predefined rules. To minimize overlaps of these deterministic functions and ensure the uniqueness of each group, we categorize the methods into the following three classes: \textit{Deterministic Graph Algorithms}, \textit{ Probabilistic Graph Algorithms}, and \textit{Heuristic-Based Graph Algorithms}. We outline the characteristics of these methods below and provide a visual representation in Figure \ref{fig:non learn alg}.

$\bullet$ \textbf{Deterministic Graph Algorithms.}  These algorithms focus on providing exact and reliable solutions to graph-related problems, ensuring correctness and precision. They prove especially effective for tasks requiring strict guarantees, such as subgraph isomorphism detection, shortest path computation, and graph traversals (e.g., Dijkstra’s algorithm \cite{dijkstra2022note}, Bellman-Ford algorithm \cite{bellman1958routing}). Their main advantage lies in their accuracy, but they may incur high computational costs for large-scale graphs. Figure \ref{fig:non learn alg} illustrates an example of the exact path computed between two nodes, highlighting how deterministic algorithms ensure optimality in path finding. 

$\bullet$ \textbf{Probabilistic Graph Algorithms.} These methods leverage probabilistic and statistical techniques to approximate graph properties or retrieve relevant information about entities and relationships. Instead of guaranteeing exact results, they offer efficient solutions for tasks where full precision is unnecessary or infeasible. Personalized PageRank, Markov Chains, and Monte Carlo methods are common examples, excelling in applications such as node ranking, influence estimation, and recommendation systems. These methods trade off determinism for scalability and adaptability in dynamic or uncertain environments. Figure \ref{fig:non learn alg} illustrates an example of a probabilistic random walk, where transitions between nodes occur with a probability 
$p$, demonstrating how these algorithms explore graph structures in a stochastic manner.

$\bullet$ \textbf{Heuristic-Based Graph Algorithms.} This category mainly includes algorithms designed to provide efficient approximate solutions, especially in cases where exact methods are computationally prohibitive. By using heuristic-based strategies or domain-specific insights, they improve scalability while maintaining reasonable accuracy. Examples include greedy algorithms for approximate subgraph matching, K-hop sampling, and subgraph sampling techniques, which are widely applied in community detection, graph partitioning, and large-scale knowledge retrieval. These methods prioritize computational feasibility and speed over absolute correctness. Figure \ref{fig:non learn alg} presents a toy example of 1-hop sampling, demonstrating how nodes efficiently gather information from their immediate neighbors.

By removing the trainable parameters in the algorithms, non-parameterized methods are relatively efficient in identifying informative subgraphs that provide valuable context for reasoning. These methods are particularly effective in scenarios where the structure of the knowledge graph is well understood, and simple traversal or ranking techniques can yield meaningful insights. Their simplicity and lack of model parameters make them easy to implement and computationally lightweight, allowing for quick and direct retrieval of relevant graph-based information to support the reasoning of LLMs.


    {
\vspace{-4mm}
\subsubsection{Methods Using Deterministic Graph Algorithms}
\textbf{GLBK~\cite{ref:noname5}} recognizes entities in the question, computes shortest paths that connect them in the knowledge graph, and retrieves text linked to the nodes and incident edges along these paths, which reveals indirect relations that support answer synthesis. \textbf{HyKGE~\cite{ref:hykge}} aligns query and hypothesis entities to a medical graph and then collects three deterministic reasoning chains between anchor entities, namely simple paths, co-ancestor chains, and co-occurrence chains, followed by pruning that keeps the most relevant connections as retrieval evidence. \textbf{GNN-RAG~\cite{ref:gnn-rag}} first scores candidate answers with graph neural reasoning and then extracts the shortest paths from the question entities to these candidates so that the retrieved paths provide faithful and efficient multi hop evidence. \textbf{KG-GPT~\cite{ref:kg-gpt}} segments the question into triple level units, matches entities to the graph through semantic alignment, identifies the relations that connect them, and assembles a subgraph that serves as compact retrieval support for downstream reasoning.
}

    {
\subsubsection{Methods Using Probabilistic Graph Algorithms}
Within the taxonomy in Table \ref{table:algorithm}, probabilistic graph algorithms cast retrieval as stochastic evidence flow on the graph. \textbf{Walklm} \cite{ref:walklm} introduces targeted random walk sampling from query anchored nodes, which turns the visited trails into textual sequences and aggregates path evidence to rank candidate neighborhoods and passages. \textbf{HippoRAG \cite{ref:hipporag}} illustrates diffusion-driven retrieval, where detected query entities are aligned to knowledge graph nodes and used as seeds for Personalized PageRank, which spreads probability through nearby structure and then applies specificity-based reweighting to select high probability neighborhoods and surface the associated passages for multi-hop reasoning.
}

    {
\vspace{-2mm}
\subsubsection{Methods Using Heuristic-Based Graph Algorithms}
\textbf{HybridRAG} \cite{ref:hybridrag} pairs vector search over the document store with a one-hop neighborhood walk from entities detected in the question, which enriches the retrieved context and improves answer quality while keeping the search path simple. \textbf{GRAG \cite{ref:grag}} illustrates structured candidate selection, where K-hop ego graphs are ranked by embedding similarity to the query. Then a learnable pruner suppresses low-value nodes and edges, and the top candidates are merged into a compact subgraph for generation, which preserves semantic focus and graph topology under tight runtime budgets.
}



\subsection{Retrieval with Learning-based Algorithms}

\subsubsection{A Unified View}

In contrast to non-parameterized approaches, the learning-based retrieval method can be formulated as a parameterized function 
$f_{\theta}(\cdot)$ that learns to retrieve relevant knowledge by optimizing an objective function:
$R_q = f_{\theta}(\mathcal{G}, e(q)),$
where $\theta$ represents the learnable parameters within 
$f_{\theta}(\cdot)$. These parameters are trained on a dataset using an optimization objective, enabling the retrieval mechanism to adapt dynamically to different data distributions and retrieval tasks. Unlike traditional rule-based retrieval, which relies on predefined heuristics, learning-based approaches can generalize across domains, capture complex semantic relationships, and improve retrieval relevance over time. Depending on how the system processes and extracts information from the knowledge graph, learning-based retrieval algorithms can be categorized into the following two major classes:

$\bullet$ \textbf{Convolutional-based Algorithms.}  Convolutional-based algorithms are designed to exploit the structural properties of graphs by performing neighborhood aggregation through trainable network layers. These methods are inspired by convolutional operations in image processing but are adapted to graphs to capture local connectivity patterns and facilitate information propagation across nodes. A representative example is the Graph Convolutional Network (GCN)~\cite{ref:gcn}, which employs a layer-wise message-passing mechanism to iteratively aggregate features from neighboring nodes. As shown on the LHS of Figure \ref{fig:non learn alg}, given the central node $x_u$ and its neighbors $x_1\sim x_3$, different weights $C_{uu}$ and $C_{u1}\sim C_{u3}$ are applied to aggregate the features from each neighbor. This weighted aggregation allows the central node to incorporate diverse contextual information, leading to richer feature representations. By stacking multiple layers, these algorithms allow retrieval models to learn hierarchical representations, enhancing their ability to extract relevant knowledge from graph-structured data.

$\bullet$ \textbf{Attention-Based Algorithms.} Attention-based algorithms in graph retrieval have gained significant interest due to their ability to dynamically prioritize the most relevant parts of the graph during learning. Unlike traditional approaches that treat all neighboring nodes equally, these methods employ attention mechanisms to assign varying importance weights ($A_{uu}$ and $A_{u1}\sim A_{u3}$ in RHS of Figure \ref{fig:non learn alg}) to nodes and edges, enabling the model to focus on key relationships that are most critical for a given task. A prominent example is the Graph Attention Network (GAT)~\cite{ref:gat}, which introduces node-wise attention coefficients to aggregate information more selectively. This adaptive weighting mechanism allows attention-based retrieval models to capture long-range dependencies, refine contextual understanding, and improve retrieval precision.


\subsubsection{Methods Using Convolutional-based Algorithms}


\textbf{REANO}~\cite{ref:reano} is a framework designed to leverage structural information in knowledge graphs to enhance the retrieval of triple features by employing an L-hop graph convolution. Before the retrieval step, REANO uses the TAGME model \cite{ferragina2010tagme} to extract entities and then builds relationships among these entities through both intra-context and inter-context manners. Subsequently, an $L$-hop GNN is applied to update entity representations, allowing REANO to select the top-$k$ triple candidates based on their similarity to the given question. By incorporating the topological structure over multiple hops within the knowledge graph, REANO gathers critical information needed to answer questions, effectively alleviating the reasoning burden of the answer predictor and improving the overall efficiency of the QA process.

\textbf{SURGE} \cite{ref:SURGE} is a framework designed to extract context-relevant knowledge from knowledge graphs while enforcing consistency across facts to ensure that the generated responses faithfully reflect the retrieved knowledge corresponding to the question. To achieve this, SURGE leverages an existing edge message passing framework \cite{jo2021edge}, which transforms the edges of the original graph into nodes within a dual hypergraph \cite{scheinerman2011fractional}. This transformation enables the use of existing node-level GNNs to represent the relationships within the original graph. Subsequently, SURGE samples negative subgraphs and employs contrastive learning, encouraging the model to generate responses that are consistent with the positive subgraph, thereby improving the fidelity of the generated content. By enforcing consistency through contrastive learning, SURGE ensures that its responses remain faithful to the retrieved knowledge, enhancing the reliability of the generated answers.


\subsubsection{Methods Using Attention-Based Algorithms}

\textbf{HSGE}~\cite{ref:hsge}~studies the complex interactions between the entities of the knowledge base by reasoning in the history semantic graph, which is built by employing a pre-trained language model
BERT \cite{ref:bert} on conversation contexts. A fundamental method of HSGE to retrieve the structure information in the history semantic graph is to update the node embeddings with the attention-based module such as TransformerConv \cite{TransformerConv}. In order to further adapt to the change triggered by the new-coming conversations, HSGE encodes the position of each historical interaction and utilizes the attention mechanism again to aggregate the most relevant information for each mentioned entity in the query. Compared with concatenating all previous conversation turns, this paradigm is more computationally efficient, while still retaining key context information needed for effective question answering.

\textbf{HamQA} \cite{ref:hamqa} intends to identify the importance of different relationships between the entities by emphasizing the information most relevant for addressing the question. Specifically, HamQA formulates the weights of each edge in the KG utilizing a learnable function combined with the entity representations. Moreover, considering the geometrically hierarchical
features between different entities, HamQA further employs  Hyperbolic distances \cite{balazevic2019multi} to measure the importance of different neighbors. These two strategies enable HamQA to effectively calculate the attention scores when incorporating the adjacent information and serve as the constraints to guide graph propagation toward more significant messages.




\begin{table}[htb]
\centering
\small
\caption{Summary of prompting methods.}
\vspace{-3mm}
\begin{tabular}{m{1.5cm}<{\centering}|m{1.8cm}<{\centering}|m{2cm}<{\centering}|m{8.4cm}<{\centering}} \toprule[1pt]
\textbf{Prompting Method} & \textbf{Pros}& \textbf{Cons}&\textbf{Reference}\\  \midrule[0.5pt]
Topology-Aware Prompting& \textit{Preserves structure, enables multi-hop reasoning} &\textit{Complex formatting, hard to understand for LLMs}&LLaGA~\cite{ref:llaga}, GNN-RAG~\cite{ref:gnn-rag}, G-Retriever~\cite{ref:g-retriever}, GRAPH-COT~\cite{ref:graphcot}, ROG~\cite{ref:rog}, RRA~\cite{ref:rra}, MVP-Tuning~\cite{ref:mvp-tuning}, Keqing~\cite{ref:keqing}, GRAG~\cite{ref:grag}, ETD~\cite{ref:etd}, DALK~\cite{ref:dalk}, MindMap~\cite{ref:mindmap}, Lark~\cite{ref:lark}, REALM~\cite{zhu2024realm}, HybridRAG~\cite{ref:hybridrag}, EWEK-QA~\cite{ref:ewek-qa}, TOG2~\cite{ref:tog2}, SURGE~\cite{ref:SURGE}, GraphGPT~\cite{ref:GraphGPT}, NLGraph~\cite{ref:nlgraph}, 
HyKGE~\cite{ref:hykge}, 
HamQA~\cite{ref:hamqa}, GraphEval2000~\cite{ref:GraphEval2000}, MuseGraph~\cite{ref:musegraph},
    {GOT~\cite{ref:got}}\\ \midrule[0.5pt]
Text Prompting &\textit{Simple implementation, compatible with LLM input format} &\textit{Loses structured information, limited reasoning ability}&Talk~\cite{ref:talk}, GPT4Graph~\cite{ref:GPT4Graph},  UniOQA~\cite{ref:unioqa}, FABULA~\cite{ref:fabula}, ODA~\cite{ref:oda}, KnowGPT~\cite{ref:knowgpt}, Kaping~\cite{ref:kaping}, MedGraphRAG~\cite{ref:medgraphrag}, Golden-Retriever~\cite{ref:golden}, KALMV~\cite{ref:kalmv}, Fact~\cite{ref:fact}, DepsRAG~\cite{ref:depsrag}, KELP~\cite{ref:kelp}, KGQA~\cite{ref:kgqa}, Graph-LLM~\cite{ref:Graph-LLM}, GLBench~\cite{ref:GLBench}, Walklm~\cite{ref:walklm}
\\ 
\bottomrule[1pt]
\end{tabular}
\vspace{-4mm}
\label{table:algorithm_prompt}
\end{table}

\subsection{Comparison between Different Retrieval Methods}
Non-parameterized and learning-based retrieval methods offer distinct advantages and trade-offs in graph-driven retrieval. Non-parameterized approaches rely on explicit graph traversal, ranking, or sampling techniques, making them computationally efficient and interpretable. They work particularly well in structured knowledge graphs, where exact or approximate methods can effectively retrieve relevant subgraphs without requiring training data. However, their performance is often limited by predefined heuristics and an inability to adapt dynamically to different retrieval contexts. In contrast, learning-based retrieval methods leverage trainable models to capture richer graph representations through message passing or attention mechanisms, allowing for more adaptive and context-aware retrieval. These methods excel at modeling complex dependencies but require substantial computational resources and labeled data for training. Moving forward, future research may focus on hybrid retrieval strategies that integrate the efficiency of non-parameterized methods with the adaptability of learning-based approaches. Additionally, improving self-supervised learning for graph retrieval can reduce dependence on labeled data, enhancing the availability in real-world applications.

\subsection{Topology-Aware Prompting}

 \subsubsection{A Unified View}

    {Topology-aware prompting captures the topological essence of a knowledge graph by explicitly encoding nodes, edges, and their relationships into structured prompts that models capable of directly processing structured inputs can readily interpret. These featured prompting methods can be divided into following categories: (1) path prompting that encodes relation chains or shortest paths to expose composition (GRAPH-COT~\cite{ref:graphcot}, ROG~\cite{ref:rog}, TOG2~\cite{ref:tog2}, MindMap~\cite{ref:mindmap}); (2) subgraph prompting that linearizes compact subgraphs for information summarization (G-Retriever~\cite{ref:g-retriever}, DALK~\cite{ref:dalk}, GRAG~\cite{ref:grag}, LLaGA~\cite{ref:llaga}); (3) rule-aware prompting that introduces relation types and declarative constraints to guide reasoning (HyKGE~\cite{ref:hykge}, HamQA~\cite{ref:hamqa}).
}

Instead of providing simple textual descriptions, these prompts explicitly represent relationships using structured formats such as triple statements (e.g., "\textit{(Albert Einstein, born in, Ulm)}") or relational paths (e.g., "\textit{Paris → capital of → France → located in → Europe}"). 
A key motivation behind graph-structure prompting is to allow the model to reason more effectively about multi-hop and complex relational patterns \cite{ref:tog2}. For instance, rather than treating the graph as a mere collection of facts, the model can examine path-based relationships to draw inferences. This is especially valuable in scenarios where the desired answer depends on understanding connections among multiple entities or interpreting intricate graph substructures. Moreover, this approach can boost explainability by making the model’s reasoning process more transparent. When each relationship or path is clearly defined, users can trace how the system derived its conclusions, thereby increasing trust and clarity in the model’s output. In summary, graph-structure prompting offers a powerful mechanism for harnessing a graph’s rich relational data, allowing RAG-based systems to deliver more reliable answers and deeper insights.

\subsubsection{Methods Applying the Topology-Aware Prompting}
\textbf{Keqing~\cite{ref:keqing}} utilizes graph-structured prompting to enhance KGQA tasks by guiding LLMs through complex multi-hop questions. It achieves this by employing two distinct prompting strategies to help LLMs understand and reason about the logical relationships in retrieved triplets from a KG. The first strategy explicitly explains the triplet structure in the prompt, detailing the format as \textit{(subject, relation, object)}, and ensuring that responses align strictly with the entities within the triplets. The second strategy converts triplets into plain text by serializing their components, making them more digestible for fine-tuned models like LLaMA 2~\cite{ref:llama}. These graph-structured prompts act as a chain-of-thought (COT) mentor for LLMs, enabling them to decompose complicated queries into simpler sub-questions, retrieve relevant entities through logical chains, and generate accurate and interpretable responses. By leveraging the structured nature of KGs and tailored prompting techniques, Keqing demonstrates improved reliability and interpretability in KGQA tasks, highlighting its potential as a scalable solution for knowledge-intensive reasoning.




Both \textbf{RoG}~\cite{ref:rog}  and \textbf{ToG}~\cite{ref:tog}  generate reasoning paths from KGs to enhance RAG prompting by structuring retrieval and reasoning. RoG focuses on generating relation-grounded paths, ensuring faithful and structured reasoning by retrieving valid reasoning paths from KGs. This approach enhances accuracy and consistency while providing interpretable results that improve trust and understanding. Similarly, ToG identifies relevant entities from a query and explores the KG by searching for meaningful triples. During reasoning, the LLM evaluates the retrieved triples and selects the most valuable ones to construct a reasoning path. This method offers explicit and editable reasoning paths, improving explainability and allowing users to trace and correct model outputs when necessary. Both approaches strengthen RAG by integrating structured retrieval with reasoning, ensuring more coherent, interpretable, and contextually relevant responses.

\subsection{Text Prompting}
\subsubsection{A Unified View}
In the cases where the machine struggles to understand the graph-structure prompt, text prompting involves transforming the structured graph knowledge with language models into a linear, textual representation that the LLMs can easily understand. After graph retrieval, the nodes, relationships, and properties are converted into descriptive sentences or paragraphs that capture the essence of the original graph while making it compatible with natural language input formats.

The focus of text prompting is on providing a human-like narrative that conveys the important details of the graph. Text prompting ensures that even complex graph structures are expressed in a format that a typical LLMs can understand, thus allowing the LLMs to utilize the graph-based information without requiring specialized graph-processing capabilities.

\subsubsection{Methods Applying the Text Prompting}

    {\textbf{MedGraphRAG} \cite{ref:medgraphrag} follows a general, reusable pattern of graph-based RAG: graph building, graph retrieval, prompting, and refinement. In the build step, an LLM reads medical documents, extracts entities, links relations, and keeps short evidence or definition snippets so that each node and edge is grounded in source text. 
 At query time, the system embeds the user question and graph entities, selects the top-k most similar entities, and expands a small working subgraph around them for focused context.
 The model then generates a draft answer from this subgraph and iteratively improves it by feeding the draft back into the LLM for a few rounds, stopping at a preset threshold and summarizing low-value regions to remain efficient. 
 This simple loop—evidence-linked graph construction, embedding-based top-k retrieval with local expansion, and short refinement—captures a general template for reliable, source-aware graph-grounded QA in high-stakes domains.}

\textbf{KCQA} \cite{ref:kgqa} builds a framework that efficiently retrieves and prompts knowledge from a KG to answer questions. Initially, KCQA estimates the distribution of relations in the KG using the Rigel model \cite{oliya2021end} and subsequently predicts relevant triples within two hops. To focus on the most relevant information, it retains only the top-K triples of the estimated relations, achieved through a Hadamard (element-wise) product to extract a weighted vector representation of the triples. These selected triples are then converted into natural language for prompting. In the prompting process, KCQA also considers inverse triples and utilizes a unified template to compose the prompt, which is then fed as the input into a language model generating an answer. 

    {
\vspace{-4mm}
\subsection{Comparison between Different Prompting Methods}
The strengths and limitations of topology-Aware prompting and text prompting map naturally to task demands. For multi-hop question answering \cite{tang2024multihop}, personalized recommendation \cite{ref:rrs}, temporal reasoning \cite{StarRAG}, and program synthesis over graphs \cite{ref:arcaneqa}, topology-aware prompting is preferable because it preserves edge types and path structure, enforces constraints, and yields auditable chains. However, as shown in Table \ref{table:algorithm_prompt}, it brings heavy formatting overhead, larger token budgets, schema canonicalization burdens, and can be harder for general LLMs to parse, making it brittle under noisy graphs \cite{hadi2023survey}. By contrast, for abstractive summarization over documents attached to the graph \cite{ref:graphrag}, for broad retrieval where lexical recall dominates \cite{ref:nutrea}, and for user-facing explanations that prioritize fluency \cite{ref:knowledgpt}, text prompting is the better choice. It is simple to implement, aligns with standard LLM inputs, remains robust under schema drift, and achieves lower latency, but it sacrifices explicit path identity, weakens constraint satisfaction, and limits compositional reasoning depth \cite{ref:gfmsurvey2}.
}
\vspace{-4mm}
\section{Graph-Structured Pipelines}\label{sec:pipeline}
\subsection{A Unified View}

\begin{wrapfigure}{r}{0.5\textwidth} 
    \centering
\vspace{-4mm}
\includegraphics[width=0.8\linewidth]{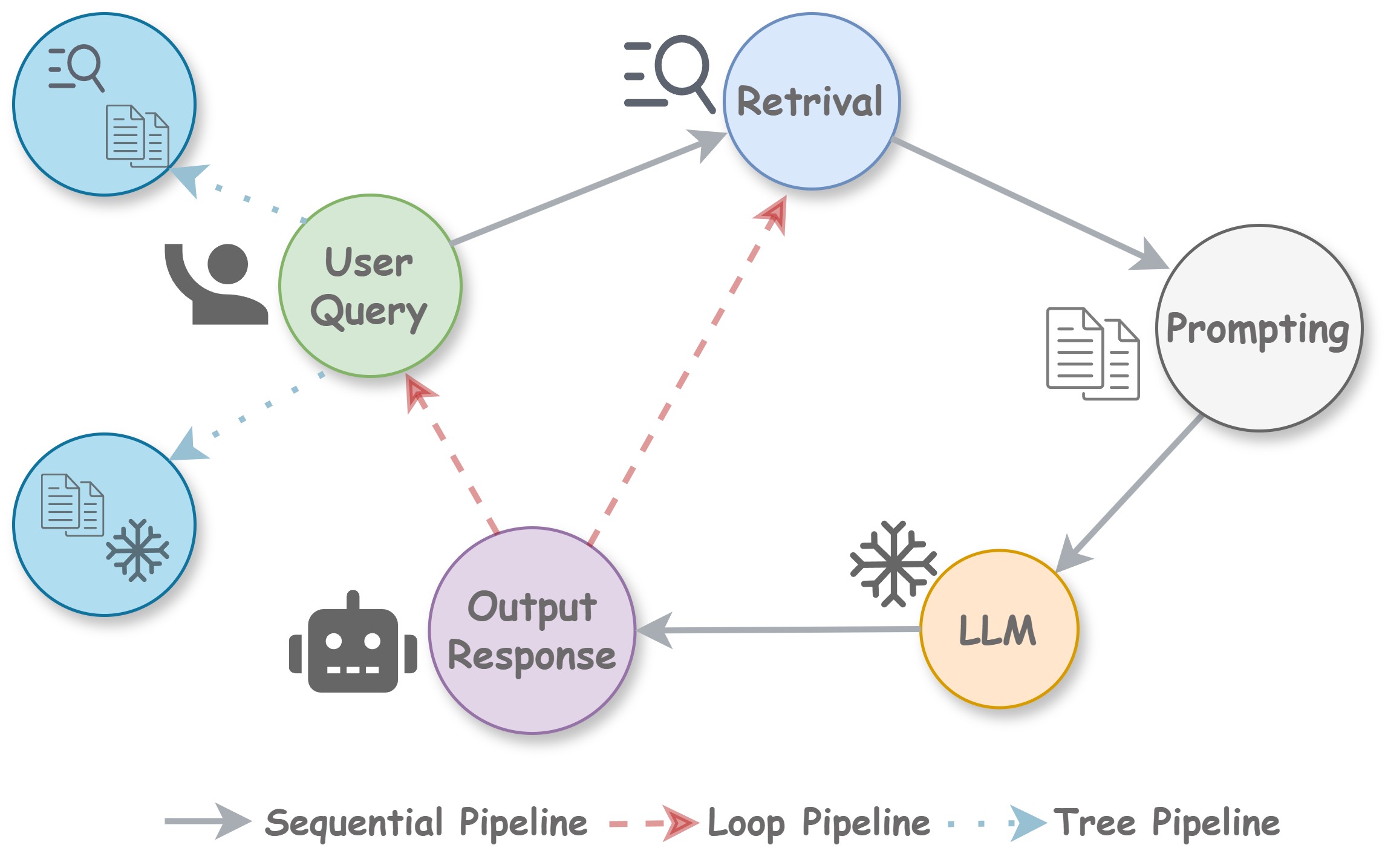}
\vspace{-4mm}
    \caption{Pipeline formulation with graph structure}
    \vspace{-4mm}
    \label{fig:pipeline}
\end{wrapfigure}
Based on the pipeline of RAG,
we can visualize the different steps in the pipeline as nodes, and the processes or transitions between them as edges. Each component in the RAG system can be represented in this graph-based structure, and the relationships between components can be defined based on their interactions. Particularly, we formulate each specific step as a node in the path of the directed graph as: (i) $e_1$: user query; (ii) $e_2$: graph retrieval; (iii) $e_3$: graph prompting; (iv) $e_4$: large language model generation; (v) $e_5$: output response. Moreover, we define the the edge $r_{i,j}$ (from node $e_i$ to node $e_j$) as: (i) $r_{1,2}$: passing the user query to the graph retrieval step; (ii) $r_{2,3}$: processing the graph retrieval output for prompting; (iii) $r_{3,4}$: passing the graph-structured or textual prompt to the LLMs; (iv) $r_{4,5}$: generating the final output response from LLMs.

In this configuration, the various steps and their transitional passages form a natural topological structure. As illustrated in Figure \ref{fig:pipeline}, we categorize the existing pipelines into three main types. (i) \textbf{Sequential Pipeline}. In reference to the figure, the sequential pipeline follows the path 
$e_1 \overset{r_{1,2}}{\longrightarrow} e_2\overset{r_{2,3}}{\longrightarrow} e_3\overset{r_{3,4}}{\longrightarrow} e_4\overset{r_{4,5}}{\longrightarrow} e_5$, mirroring the standard RAG paradigm shown in Figure \ref{fig:rag}. This approach represents a linear flow, wherein each step immediately succeeds the previous one, and no node is revisited.
(ii) \textbf{Loop Pipeline}. In this variant, certain stages incorporate feedback mechanisms or iterative processes. For instance, the graph retrieval step may be repeated if the initial retrieval fails to supply adequate context, or the LLMs might require additional prompts to refine their responses. We capture this behavior with a cyclic graph. In Figure \ref{fig:pipeline}, the final output can also be routed back to earlier steps, either merging with the user query ($e_5\overset{r_{5,1}}{\longrightarrow} e_1$) or informing another retrieval phase ($e_5\overset{r_{5,2}}{\longrightarrow} e_2$).
(iii) \textbf{Tree Pipeline}. This paradigm is designed for scenarios in which multiple components of the pipeline can execute in parallel. The graph splits into multiple branches, allowing different retrieval or prompting strategies to occur simultaneously. In the following sections, we introduce representative methods employing these different pipeline designs and summarize a widely investigated classification in Table \ref{table:pipeline}.

\vspace{-2mm}
\begin{table}[htb]
\centering
\small
\caption{Summary of graph-structured pipelines.}
\vspace{-2mm}
\begin{tabular}{m{1.5cm}<{\centering}|m{1.5 cm}<{\centering}|m{1.5 cm}<{\centering}|m{8.8cm}<{\centering}} \toprule[1pt]
\textbf{Pipelines}& \textbf{Pros}& \textbf{Cons} & \textbf{Reference}\\  \midrule[0.5pt]
Sequential Pipeline&\textit{Simple and efficient execution} &\textit{Fixed order, no refinement or error correction}& GNN-RAG~\cite{ref:gnn-rag}, GrapeQA~\cite{ref:grapeqa}, SR~\cite{ref:sr}, RRA~\cite{ref:rra}, KG-GPT~\cite{ref:kg-gpt}, QA-GNN~\cite{ref:qa-gnn}, GLBK~\cite{ref:noname5}, ATLANTIC~\cite{ref:atlantic}, UniOQA~\cite{ref:unioqa}, GRAG~\cite{ref:grag}, HyKGE, ~\cite{ref:hykge}, KG-Rank~\cite{ref:kg-rank}, KnowledGPT~\cite{ref:knowledgpt}, ETD~\cite{ref:etd}, HippoRAG~\cite{ref:hipporag}, DALK~\cite{ref:dalk},  KnowGPT~\cite{ref:knowgpt}, Engine~\cite{ref:engine}, GraphBridge~\cite{ref:graphbridge}, ChatKBQA~\cite{ref:chatkbqa}, NuTrea~\cite{ref:nutrea}, etc  \\ \midrule[0.5pt]
Loop Pipeline & \textit{Iterative refinement, error correction}&\textit{Higher computational cost, low efficiency}&GraphRAG~\cite{ref:graphrag}, GRAPH-COT~\cite{ref:graphcot}, KnowledgeNavigator~\cite{ref:kn}, KD-COT~\cite{ref:kd-cot}, KG-Agent~\cite{ref:kg-agent}, GNN-Ret~\cite{ref:gnn-ret}, Knowledge Solver~\cite{ref:ksl}, KGP~\cite{ref:kgp}, FABULA~\cite{ref:fabula}, ODA~\cite{ref:oda}, QR-LLM~\cite{ref:rewrite1}, MufassirQAS~\cite{alan2024rag}, ArcaneQA~\cite{ref:arcaneqa}, MedGraphRAG~\cite{ref:medgraphrag}, KALMV~\cite{ref:kalmv}
\\ \midrule[0.5pt]
Tree Pipeline & \textit{Parallel exploration, multi-path reasoning}& \textit{Higher computational cost}& EWEK-QA~\cite{ref:ewek-qa}, HamQA~\cite{ref:hamqa}, SURGE~\cite{ref:SURGE}, TOG2~\cite{ref:tog2},
  {GOT~\cite{ref:got}},
  {Beyond-CoT~\cite{ref:Beyond_CoT}}\\
\bottomrule[1pt]
\end{tabular}
\vspace{-5mm}
\label{table:pipeline}
\end{table}

\subsection{Controllers of Pipeline Implementation}
The \emph{controller} is the decision-making mechanism that formalizes our pipeline topologies at inference time, which chooses when to retrieve, whether to iterate or branch, when to verify and revise, and when to stop. Different controller families emphasize sequential, iterative loop, and branching tree pipelines, each with distinct tradeoffs among accuracy, latency, and computational cost. In what follows, we present four widely adopted controller families in RAG systems and show how representative methods instantiate each family within our pipeline design.

\textbf{Reinforcement-Learning Controllers.} A reinforcement-learning (RL) controller is a learned policy that, given the current reasoning state, selects the next pipeline action (retrieve, expand, verify, revise, or stop) to maximize a reward that trades off answer quality and computational cost. For example, {RAG-RL} \cite{ref:Rag-rl} post-trains the generator with RL so it learns to select and cite useful contexts from larger retrieved sets, which tightens the coupling between retrieval and generation and effectively learns the loop and tree branching rules for exploration and pruning. 
 Multi-module RL has also been used to coordinate several RAG components jointly (query rewriter, retriever, selector, generator), as in {MMOA-RAG} \cite{ref:MMOA-RAG}, which frames the pipeline as cooperative agents optimized toward a unified reward and thus implements a learned loop controller and branch selection over candidate contexts.

 \textbf{System-2 Slow-Thinking Controllers.} A system-2 controller is an inference-time search policy that allocates more computation than a single forward pass, which naturally realizes loop and tree pipelines without updating model weights. For instance, {Self-Ask} \cite{ref:self-ask} decomposes a question into sub-questions, interleaves retrieval during the reasoning process, and thus implements a loop that alternates “ask–retrieve–answer” over a constrained subgraph. 
 {FLARE} \cite{ref:FLARE} generalizes this idea to long-form generation by forecasting upcoming content, triggering retrieval only when predicted segments are low confidence, and regenerating with fresh evidence.

\textbf{Verifier-Guided Controllers}. A verifier-guided controller attaches an external evaluator that scores partial solutions or retrieved sets and feeds decisions back to the pipeline, thereby governing revision and selection. {Self-RAG} \cite{ref:Self-rag} trains a model to retrieve, generate, and \emph{critique} using reflection tokens. During inference, it adaptively decides whether to retrieve again, accept or revise a segment, and produce citations, which instantiates a loop controller with verify–revise cycles over the current subgraph. 
{CRAG} \cite{CRAG} adds a lightweight retrieval evaluator that estimates the quality of retrieved documents and triggers corrective actions, such as re-retrieval or web fallback, effectively steering the loop and pruning poor branches in a tree of candidate evidence.

\textbf{Budget-/Cost-Aware Controllers.} A budget-aware controller explicitly optimizes an accuracy–latency objective by selecting per-query computation and retrieval budgets, thereby choosing whether a sequence suffices or whether a loop or tree is warranted.{Adaptive-RAG} \cite{ref:Adaptive} predicts query complexity and selects among no-retrieval, single-step retrieval, or iterative retrieval strategies, implementing a controller that upgrades a sequence pipeline into a loop only when complexity demands. 
 Conversational gating methods such as {RAGate} \cite{ref:RAGate} estimate, at each turn, whether external augmentation is needed and hence decide to skip or invoke retrieval, providing a cost-aware switch between sequence and loop behavior under compute or latency constraints.

\subsection{Methods Applying the Sequential Pipeline}

\textbf{Keqing \cite{ref:keqing}}  adopts a sequential pipeline to address complex multi-hop question answering.  It decomposes complex queries into simpler sub-questions using predefined templates. These sub-questions are then aligned with logical chains on a knowledge graph (KG), guiding the retrieval of candidate entities through multi-hop reasoning. A candidate reasoning module evaluates and selects the correct entities from the KG. Finally, the system generates responses by aggregating reasoning paths and answers to provide an interpretable and precise answer to multi-hop questions.

\textbf{QA-GNN \cite{ref:qa-gnn}} integrates LMs and KGs for question answering with a sequential pipeline. The process begins by encoding the QA context including the question and answer options with an LM and retrieving a KG subgraph relevant to the question. A joint "working graph" is constructed by connecting the QA context node to entities within the KG subgraph. To enhance reasoning, the KG nodes are scored for relevance based on their alignment with the QA context using LM-based scoring. An attention-based GNN performs iterative message passing on this working graph, updating both KG entities and QA context representations. Finally, predictions are made by aggregating the LM output, updated node features, and the working graph representation, achieving accurate and explainable results.

\textbf{DALK \cite{ref:dalk}}  makes LLMs and KGs benefit from each other through an elaborate sequential pipeline design. At the beginning of the pipeline, LLMs process the scientific corpus and construct two domain-specific KGs from the related literature via pair-wised relation extraction and generative relation extraction, respectively. This part of the pipeline can be seen as LLMs benefit KGs. Then, DALK utilizes a coarse-to-fine sampling method and a retrieval approach to select knowledge from KGs. The selected knowledge as well as the domain questions are sent to LLMs to generate accurate answers, therefore realizing that KGs benefit LLMs.









\vspace{-2mm}
\subsection{Methods Applying the Loop Pipeline}





\textbf{RGNN-Ret \cite{ref:gnn-ret}} tackles multi-hop reasoning questions by iteratively decomposing the reasoning process into manageable steps. It incorporates a self-critique mechanism that prompts LLMs to generate sub-questions for each reasoning step. After answering a sub-question with retrieved passages, the LLM evaluates whether the accumulated evidence is sufficient to generate a final answer or if further reasoning steps are required. This iterative process enables the system to dynamically adapt to the complexity of the query.
To enhance the retrieval quality across reasoning steps, a Recurrent Graph Neural Network (RGNN) is used. The RGNN integrates the graphs of passages from previous reasoning steps, establishing connections between retrieved passages to ensure that relationships between sub-questions and evidence are maintained. By combining semantic distances and contextual information across steps, the RGNN reduces the impact of incorrect sub-questions and improves retrieval accuracy for supporting passages.
The iterative loop continues until the self-critique determines that enough evidence has been collected to answer the question.




 \textbf{GRAPH-COT \cite{ref:graphcot}}  provides a benchmark to enhance LLM reasoning on graphs through iterative steps of reasoning, interaction, and execution. At each iteration, the LLM identifies the required information and formulates graph interactions to retrieve relevant data, such as nodes or relationships. These interactions are executed on the graph, and the results are fed back into the reasoning process. This cycle repeats until the LLM concludes the reasoning task and produces the final answer. By iteratively refining its understanding of graph structures, GRAPH-COT effectively handles multi-hop reasoning tasks, ensuring accurate and explainable results in graph-based question answering.


\vspace{-2mm}
\subsection{Methods Applying the Tree Pipeline}

\textbf{SURGE} \cite{ref:SURGE} effectively leverages a tree pipeline by structuring its retrieval and generation processes into multiple parallel branches, allowing different retrieval and prompting strategies to execute simultaneously. Instead of sequentially retrieving a single set of facts, SURGE retrieves multiple context-relevant subgraphs in parallel, each representing different aspects of the dialogue history. These subgraphs are processed simultaneously through distinct embedding pathways, ensuring diverse knowledge representations. The model then applies its invariant graph encoding technique to each retrieved subgraph independently, maintaining permutation and relation-inversion invariance across multiple branches. Additionally, during response generation, SURGE integrates these parallel knowledge streams using graph-text contrastive learning, ensuring that the generated response remains consistent across all retrieved knowledge sources. This parallel execution not only enhances efficiency but also enables the model to synthesize information from multiple perspectives, improving response accuracy and informativeness in knowledge-grounded dialogue systems.

  {\textbf{GoT} \cite{ref:got} effectively leverages a tree pipeline by expanding an initial prompt into multiple parallel branches of “thoughts,” each branch pursuing a distinct partial solution under a predefined Graph of Operations. Rather than committing to a single chain, GoT repeatedly applies a generate–score–select routine: candidate branches are spawned with a controllable branching factor, locally validated and scored, and then pruned via a keep-best policy to preserve only the most promising subtrees. This staged branching decomposes the task into tractable subproblems that can be solved concurrently within separate branches, while refinement loops allow individual branches to self-correct without restarting the whole pipeline. Crucially, GoT then fuses the surviving branches using explicit aggregation operators, turning the pure tree into a directed acyclic workflow that consolidates partial results into a coherent final output. By combining parallel expansion, principled pruning, iterative refinement, and late-stage aggregation, GoT maintains the efficiency benefits of a tree pipeline across the entire process while achieving higher solution quality and better cost–latency trade-offs than sequential methods.}

  {\textbf{Beyond-CoT \cite{ref:Beyond_CoT}} leverages a tree-style pipeline by expanding a single query into multiple concurrent branches that persist across two stages—rationale generation and answer inference. First, an Extract–Cluster–Coreference (ECC) routine builds a branched thought graph from the input and optional image caption, effectively fanning out the problem into many node-level sub-traces. Each branch is then encoded in parallel: text via a Transformer encoder, the thought graph via a graph attention network, and vision via a dedicated encoder. Cross-attention aligns tokens with graph nodes, and a gated-fusion layer aggregates these parallel branches into a unified representation that the decoder uses to produce rationales. In stage two, the pipeline re-branches by concatenating the predicted rationales back to the input, reconstructing the thought graph, and repeating the same parallel encoders, alignment, and gated fusion before decoding the final answer. This end-to-end design preserves the efficiency and diversity benefits of a tree pipeline—branching for exploration, parallel encoding for coverage, and late fusion for consolidation—while generalizing the tree into a graph to capture non-linear dependencies.}

\subsection{Comparison between Different Pipelines}
Different pipelines offer varying degrees of flexibility and efficiency, influencing how information flows through RAG systems. Sequential pipelines provide a straightforward, stepwise execution, ensuring clarity and simplicity but limiting adaptability when intermediate refinements are needed. In contrast, loop pipelines introduce feedback mechanisms, allowing iterative refinement of retrieved knowledge or prompt modifications, making them well-suited for tasks requiring multiple rounds of reasoning. However, they can introduce computational overhead due to repeated processing. Tree pipelines, on the other hand, enable parallel execution of different retrieval or prompting strategies, improving efficiency in handling diverse queries but potentially increasing system complexity. A future trend of pipelines is moving toward adaptive pipeline selection, where the system dynamically determines the optimal pipeline structure based on the query’s complexity and reasoning needs \cite{jiang2025d}. This strategy can enable more efficient, context-aware retrieval and generation while minimizing unnecessary computational overhead.

\section{Graph-Oriented Tasks}\label{sec:task}
\subsection{KGQA Tasks}
\subsubsection{A Unified View}
Knowledge Graph Question Answering (KGQA) targets answering natural language queries by leveraging the structured information encoded in KGs. Unlike traditional question-answering methods that rely solely on text-based retrieval, KGQA focuses on grounding its reasoning in the relationships and entities represented within a knowledge graph. By utilizing the rich semantic structure of KGs, which capture entities, their attributes, and the interrelationships between them, KGQA systems aim to provide precise and contextually accurate answers. This task not only requires understanding the natural language query but also necessitates navigating the graph’s structure to extract relevant facts, perform logical inferences, and deliver a well-supported response. In the ensuing section, we offer in-depth discussions of select noteworthy works, and a more extensive overview of additional works can be found in Figure \ref{table:task}.  {To better summarize these methods, we leave the brief introduction of them in the Appendix.}

 \subsubsection{Methods Targeting the KGQA Tasks}
\textbf{G-Retriever \cite{ref:g-retriever}} is designed to improve KGQA by addressing challenges in reasoning, scalability, and accuracy for real-world textual graphs. It introduces a new benchmark of graph-based question answering, enabling models to handle complex, multi-domain questions beyond basic graph reasoning tasks. A key innovation of this benchmark is formulating the retrieval of subgraph as a Prize-Collecting Steiner Tree optimization \cite{bienstock1993note} problem, ensuring relevant information is extracted while maintaining explainability. G-Retriever integrates GNNs and LLMs for fine-tuned reasoning and achieves superior performance across domains such as knowledge graphs, scene graphs, and common-sense reasoning. By offering a conversational interface for graph queries, it advances the explainability, efficiency, and usability of KGQA tasks. This makes it a benchmark in advancing graph-based question-answering systems.

\textbf{GraphRAG \cite{ref:graphrag}} enhances KGQA by integrating traditional RAG with graph-based summarization, addressing global queries that traditional RAG struggles to resolve. It builds an LLM-derived knowledge graph that uses community detection algorithms, such as Leiden, to group related entities and relationships into modular communities. These communities are pre-summarized, allowing for efficient multi-hop reasoning and improved retrieval accuracy. For answering queries, partial responses from community summaries are combined into a global answer through a map-reduce approach. This method improves comprehensiveness, diversity, and scalability, enabling precise, contextually rich responses for large-scale datasets with significantly reduced computational overhead.






\begin{table}[htb]
\centering
\small
\caption{Summary of different tasks. The brief introduction of the listed methods can be found in the appendix.}
\vspace{-3mm}
\begin{tabular}{m{3cm}<{\centering}|m{11cm}<{\centering}} \toprule[1pt]
\textbf{Tasks} & \textbf{Reference}\\  \midrule[0.5pt]
KGQA Tasks&
HSGE~\cite{ref:hsge}, ReTraCk~\cite{ref:retrack}, RNG-KBQA~\cite{ref:rng}, ArcaneQA~\cite{ref:arcaneqa}, EWEK-QA~\cite{ref:ewek-qa}, HamQA~\cite{ref:hamqa}, KG-FiD~\cite{ref:kgfid}, Golden-Retriever~\cite{ref:golden}, SURGE~\cite{ref:SURGE},KELP~\cite{ref:kelp}, Fact~\cite{ref:fact}, TOG2~\cite{ref:tog2}, MINERVA~\cite{ref:minerva}, KALMV~\cite{ref:kalmv},REANO~\cite{ref:reano},LLaGA~\cite{ref:llaga}, GNN-RAG~\cite{ref:gnn-rag}, G-Retriever~\cite{ref:g-retriever}, GraphRAG~\cite{ref:graphrag},  ROG~\cite{ref:rog}, SR~\cite{ref:sr}, RRA~\cite{ref:rra}, KnowledgeNavigator~\cite{ref:kn}, KG-GPT~\cite{ref:kg-gpt},  KD-COT~\cite{ref:kd-cot}, QA-GNN~\cite{ref:qa-gnn},  MVP-Tuning~\cite{ref:mvp-tuning}, StructGPT~\cite{ref:structgpt}, MHKG~\cite{ref:feng}, MHGRN~\cite{ref:noname3}, Temple-MQA~\cite{ref:temple-mqa}, KagNet~\cite{ref:kagnet}, KG-Agent~\cite{ref:kg-agent}, GenKGQA~\cite{ref:gentkgqa}, GLBK~\cite{ref:noname5},  Keqing~\cite{ref:keqing}, UniKGQA~\cite{ref:unikgqa},  GRAG~\cite{ref:grag},
Knowledge Solver~\cite{ref:ksl},
WebQSQ~\cite{ref:webqsp},
MetaQA~\cite{ref:metaqa},
PullNet~\cite{ref:pullnet},
KnowledGPT~\cite{ref:knowledgpt},
ETD~\cite{ref:etd},
DecAF~\cite{ref:decaf},HippoRAG~\cite{ref:hipporag}, GGE~\cite{ref:gge}, NSM~\cite{ref:nsm}, SKP~\cite{ref:skp}, KnowGPT~\cite{ref:knowgpt}, Difar~\cite{ref:difar}, Kaping~\cite{ref:kaping}, NuTrea~\cite{ref:nutrea}, OreoLM~\cite{ref:oreolm}
\\ \midrule[0.5pt]
Graph Tasks & GPT4Graph~\cite{ref:GPT4Graph}, GraphText~\cite{ref:graphtext},  LLaGA~\cite{ref:llaga}, GRAPH-COT~\cite{ref:graphcot}, Engine~\cite{ref:engine}, GraphBridge~\cite{ref:graphbridge},  MMGCN~\cite{ref:mmgcn},ConvE~\cite{ref:wn}, Graph-LLM~\cite{ref:Graph-LLM}, GraphGPT~\cite{ref:GraphGPT}, NLGraph~\cite{ref:nlgraph}, GLBench~\cite{ref:GLBench}, GraphEval2000~\cite{ref:GraphEval2000}, MuseGraph~\cite{ref:musegraph},  Walklm~\cite{ref:walklm} 
\\ \midrule[0.5pt]
Domain-specific Tasks & GLBK~\cite{ref:noname5}, ATLANTIC~\cite{ref:atlantic}, HyKGE~\cite{ref:hykge}, KG-Rank~\cite{ref:kg-rank},  DALK~\cite{ref:dalk}, KGP~\cite{ref:kgp},  MindMap~\cite{ref:mindmap}, FABULA~\cite{ref:fabula}, FoodGPT~\cite{ref:foodgpt}, KAM-CoT~\cite{ref:kamcot}, MufassirQAS~\cite{alan2024rag}, REALM~\cite{zhu2024realm}, MEDQA~\cite{ref:medqa}, HybridRAG~\cite{ref:hybridrag}, MedGraphRAG~\cite{ref:medgraphrag}, DepsRAG~\cite{ref:depsrag} \\
\bottomrule[1pt]
\end{tabular}
\vspace{-2mm}
\label{table:task}
\end{table}

\vspace{-2mm}
\subsection{Graph Tasks}
\subsubsection{A Unified View}
By incorporating LLMs into graph task execution, the retrieval of relevant subgraphs or neighboring nodes can improve the accuracy of tasks like node classification, link prediction, and graph classification. These tasks benefit from the reasoning capabilities of LLMs, which can understand complex relationships within the graph. Specifically, by utilizing the LLM’s ability to process and reason over graph structures, predictions become more accurate, allowing for better insights and performance across a variety of graph-based applications. 

 \subsubsection{Methods Targeting the Graph-centric Tasks}
\textbf{RAGRAPH}~\cite{ref:ragraph} performs node classification, link prediction, and graph classification tasks.
By retrieving and leveraging the most relevant subgraphs, RAGRAPH passes information within subgraphs to enhance the representation of the center node, facilitating various graph learning tasks. \textbf{LitFM}~\cite{ref:litfm} considers link prediction (citation link prediction, paper recommendation) and text generation (title generation, abstract completion, citation sentence generation) tasks.

\textbf{GraphGPT} \cite{ref:GraphGPT} aims to align LLMs with graph learning tasks through a graph instruction tuning paradigm. The framework focuses on integrating graph structural knowledge with LLMs to enhance generalization in both zero-shot and supervised settings. GraphGPT includes a text-graph grounding component that aligns graph structures with natural language, allowing LLMs to comprehend complex graph structures. This alignment is achieved through a dual-stage instruction tuning process: first, self-supervised instruction tuning is employed to provide structural knowledge, and second, task-specific instruction tuning is used to enhance adaptability across various graph tasks, including node classification and link prediction. GraphGPT also leverages COT distillation to improve step-by-step reasoning abilities, making it effective at addressing the challenges of zero-shot learning in graph-based tasks.

\subsection{Domain-specific Tasks}
GraphRAG can be effectively employed in diverse domains such as academia, e-commerce, scientific literature, healthcare, and legislation due to its ability to integrate powerful language generation models with retrieval-based knowledge. 

 \textbf{Healthcare.} Existing literature leverages RAG to efficiently summarize long-form medical documents, enhancing LLMs' ability to generate accurate and evidence-based responses. For instance, {MedGraphRAG}~\cite{ref:medgraphrag} serves as a dedicated framework for utilizing graph-based RAG in healthcare applications. Its retrieval process involves generating a tag-summary on the user query, identifying the most relevant graph through similarity-based selection, and refining responses. {HyKGE}~\cite{ref:hykge} is a framework leveraging RAG based on knowledge graphs in LLM-empowered medical applications. 
 To improve medical consultation quality, HyKGE integrates a granularity-aware reranking module to eliminate noise while preserving diversity-relevance balance in retrieved knowledge.
 {REALM}~\cite{zhu2024realm} enhances the clinical predictive capabilities by integrating RAG with multimodal Electronic Health Records (EHR). It enhances the utilization of EHR data in healthcare and reconciles it with the intricate medical context necessary for educated clinical forecasts. 
    
    \textbf{Scientific Literature.} By harnessing the capabilities of GraphRAG, it becomes possible to navigate the vast expanse of scientific literature, extracting key insights and condensing complex ideas and high-quality knowledge. For example, {DALK}~\cite{ref:dalk}  constructs a specific knowledge graph sourced from Alzheimer's Disease scientific literature. The essential knowledge is filtered through a coarse-to-fine sampling algorithm so that it can offer valuable insights. {ATLANTIC}~\cite{ref:atlantic} introduces a retrieval-augmented language model that incorporates structural awareness for interdisciplinary scientific tasks, integrating heterogeneous document graphs to capture structural relationships among scientific documents. By fusing textual and structural embeddings, ATLANTIC enhances the retrieval of coherent and faithful passages. {KGP}~\cite{ref:kgp}  creates a KG over multiple kinds of literature and employs an LLM-guided traversal agent to retrieve and synthesize relevant information for answering complex queries.

     \textbf{Code Completion.} In code completion tasks, using a graph with the RAG system helps capture and represent the complex relationships between code elements, such as control flow, data dependencies, and function calls. For example, {GraphCoder} \cite{ref:graphcoder} is an RAG framework designed to improve code completion by integrating repository-specific knowledge into code LLMs. Unlike traditional methods, it uses a Code Context Graph (CCG), which captures relationships like control flow and data dependence. The framework employs a two-step retrieval process: coarse-grained filtering to find candidate code snippets, followed by fine-grained re-ranking to prioritize those with aligned dependencies. This combination of structural and lexical context leads to more relevant code snippets, improving the accuracy of code generation. GraphCoder outperforms baseline methods, achieving higher code and identifier exact matches with lower time and storage costs. Its language-agnostic design, tested on Python and Java, proves its versatility across different programming environments.

 \textbf{Biomedical Question Answering.} Biomedical knowledge is specialized and constantly changing, making it hard for traditional language models to provide accurate, up-to-date answers without access to external, structured information.
{GLBK \cite{ref:noname5}} aims to tackle the information overload problem in biomedical question answering. The primary objective is to retrieve and prioritize relevant biomedical documents from a vast corpus, such as PubMed, while addressing complex, open-ended queries like identifying drug targets for diseases. Unlike traditional methods relying solely on embedding similarity, this approach integrates knowledge graph structures to mitigate biases from overrepresented topics and improve access to the long tail of biomedical knowledge. By leveraging entity recognition, relationship extraction, and graph-based rebalancing, the method enhances retrieval precision and ensures more contextually relevant information is surfaced. 

\vspace{-3mm}
 {\subsection{Evaluation of Tasks}
\textbf{Open-source benchmark datasets.} Following the taxonomy above, we also group the open-source benchmark datasets into three representative families (summarized in Table~\ref{table:evaluation}). For \emph{KGQA}, WebQ \cite{ref:webQ} and WebQSP \cite{ref:webqsp} focus on entity linking and compositional queries over knowledge graphs, while HotpotQA \cite{ref:hotpotqa} emphasizes multi-hop aggregation and GrailQA \cite{ref:grailqa} targets schema generalization across unseen relations. Recent temporal KGQA corpora—CronQuestion \cite{ref:cronquestion}, Forecast \cite{ref:forecast}, and MultiTQ \cite{ref:multitq}—introduce explicit time constraints to test era-aware retrieval and timestamp-valid answers. For \emph{graph-centric tasks}, RelBench \cite{ref:relbench} provides standardized node, edge, and graph prediction suites. TUDataset \cite{ref:tudataset} offers diverse graph classification collections, and GLBench \cite{ref:GLBench} together with GraphEval2000 \cite{ref:GraphEval2000} concentrate on LLM-oriented graph reasoning and program execution. In \emph{domain-specific settings}, MedQA \cite{ref:medqa} and PubMedQA \cite{ref:pubmedqa} address clinically grounded QA with domain terminology and long-context evidence. BioRED \cite{ref:biored} targets biomedical entity and relation extraction, and AlphaFin \cite{li2024alphafin} represents financial analysis and forecasting under realistic market dynamics. }

\begin{table}[htb]
\centering
\small
\vspace{-3mm}
\caption{ {Evaluation of different tasks.}}

\vspace{-2mm}
\begin{tabular}{m{2cm}<{\centering}|m{4 cm}<{\centering}|m{8 cm}<{\centering}} \toprule[1pt]
\textbf{Tasks}& \textbf{Metrics} & \textbf{Benchmark datasets}\\  \midrule[0.5pt]
KGQA Tasks& Exact Match, F1 Score, SCR, Accuracy, ROUGE, BLEU, METEOR, TTFT  &  SimpleQuestion \cite{ref:simplequestion}, WebQ \cite{ref:webQ}, CWQ \cite{ref:cwq}, MetaQA \cite{ref:metaqa}, {CronQuestion} \cite{ref:cronquestion}, {Forecast} \cite{ref:forecast}, {MultiTQ} \cite{ref:multitq}, GraphQA \cite{ref:g-retriever}, GRBENCH \cite{ref:graphcot}, CRAG \cite{ref:crag}, NQ \cite{ref:nq}, KILT \cite{ref:zsre}, MultiHop \cite{tang2024multihop}, CRUD-RAG \cite{CRUD}, CREAK \cite{ref:creak}, HotpotQA \cite{ref:hotpotqa}, TriviaQA \cite{ref:triviaqa}, FACTKG \cite{ref:factkg}, WebQSP \cite{ref:webqsp}, GrailQA \cite{ref:grailqa}, 
\\ \midrule[0.5pt]

Graph Tasks &Accuracy, F1 Score, MRR, NDCG, Hit
Ratio, MAE, RMSE  & RelBench \cite{ref:relbench}, TabGraphs \cite{ref:tabgraphs}, HiTab \cite{ref:hitab}, TUDataset \cite{ref:tudataset}, GLBench \cite{ref:GLBench}, GPT4Graph \cite{ref:GPT4Graph}, CS-TAG \cite{ref:CS-TAG}, GraphEval2000 \cite{ref:GraphEval2000}
\\ \midrule[0.5pt]
Domain-specific Tasks & Accuracy, F1 Score, MRR, ROUGE, RMSD, Cumulative Return & STARK \cite{ref:stark}, AlphaFin \cite{li2024alphafin}, MedQA \cite{ref:medqa}, BioRED \cite{ref:biored}, QALD \cite{ref:qald}, Mintaka \cite{ref:mintaka}, TutorQA \cite{ref:Graphusion}, UltraDomain \cite{ref:Memorag}, PubMedQA \cite{ref:pubmedqa}
\\\bottomrule[1pt]
\end{tabular}
\vspace{-3mm}
\label{table:evaluation}
\end{table}

 {
\textbf{Evaluation metrics.} We summarize widely used metrics across the three task families and note that Accuracy and F1 score are ubiquitous core measures for them. For answer quality in, Exact Match (EM), token-level F1 score, and Accuracy are standard, with Recall-Oriented Understudy for Gisting Evaluation (ROUGE), Bilingual Evaluation Understudy (BLEU), and Metric for Evaluation of Translation with Explicit ORdering (METEOR) applied when systems produce long-form justifications. For ranking and retrieval in graph reasoning, Mean Reciprocal Rank (MRR), Normalized Discounted Cumulative Gain (NDCG), and Hits@k are additionally reported. In domain-specific tasks, biomedical information extraction typically uses Accuracy and micro or macro F1; biology-oriented structure comparison relies on Root Mean Square Deviation (RMSD); and finance-oriented studies commonly track Cumulative Return, Sharpe Ratio, Maximum Drawdown and so on. Efficiency and stability are often captured with latency measures such as Time to First Token (TTFT) and robustness measures such as Self-Consistency Rate (SCR). The complete introduction of each metric is detailed in the appendix.
}

\vspace{-3mm}
 {
\subsection{Importance of Graph in Different Task Families}
We finally examine the advantages of graph-based RAG over non-graph retrieval across three task families. 
(a) For simple factoid KGQA, a non-graph retriever often suffices because the answer typically resides in a single passage with little need for disambiguation. For complex KGQA, graphs add clear value by grounding entities and relations, composing multi-hop paths before generation, enforcing type and temporal constraints, constraining search to relevant neighborhoods, and preserving provenance for verification. These capabilities are instantiated by the Prize Collecting Steiner Tree module in G-Retriever, which selects an evidence subgraph aligned to the question, and the Personalized PageRank diffusion module in HippoRAG, which prioritizes neighborhoods seeded by detected entities to guide ranking and multi-hop reasoning. (b) For graph-centric tasks such as node classification and link prediction, performance hinges on local topology, relation types, and neighborhood signals. Graph-based RAG contributes by retrieving a compact subgraph anchored on the target nodes or node pairs, preserving adjacency and type constraints so the generator reasons over dense, topology-aware evidence within a small token budget. This reduces distraction from long passages, improves label and edge consistency, and supports clear provenance. These advantages are exemplified by subgraph message passing modules in RAGRAPH, which strengthen center node representations for node and link prediction, and the feature\&text level enhancement modules in LLMs- EP, which inject two-hop structural signals to improve node classification. (c) Domain-specific tasks demand high accuracy and transparent explanations. Graph-based RAG strengthens accuracy by supporting loop and tree pipelines that let the model revisit, branch, and refine retrieval and generation across multiple passes. It also advances explainability by representing structured knowledge beyond classic knowledge graphs, including document, citation, clinical, and code graphs, so the system returns subgraphs with explicit entities, relations, and paths as verifiable evidence. These advantages are illustrated by the traversal agent in KGP, which iteratively expands and consolidates evidence on literature graphs, and the shortest-path retrieval in GLBK, which surfaces indirect biomedical relations with clear provenance.
}

\vspace{-3mm}
 {
\section{Graph-based RAG in dynamic scenarios}
Graph-based RAG in dynamic scenarios increasingly couples time-aware indexing with retrieval that respects event order and temporal constraints. T-GRAG \cite{ref:T-GRAG} constructs a time-stamped knowledge graph and performs layered retrieval that first isolates time-relevant subgraphs, then filters candidate nodes and finally selects fine-grained facts for generation, thereby avoiding cross-time conflicts. DyG-RAG \cite{ref:DyG-RAG} defines dynamic event units with explicit temporal anchors and links them into an event graph, retrieving coherent timelines and guiding generation with time-aware chain-of-thought. STAR-RAG \cite{StarRAG} compresses temporal regularities into a rule graph and applies seeded personalized PageRank to prioritize a time-aligned neighborhood that matches a query’s temporal intent. TimeRAG \cite{ref:timeRAG} builds a time series knowledge base and retrieves reference sequences via Dynamic Time Warping before prompting an LLM forecaster with the retrieved patterns. TS-Retriever \cite{ref:TS-retriever} improves retrieval for time-sensitive questions by training a contrastive, temporally aware retriever with query-side fine-tuning and routing, so that queries with implicit or interval constraints are matched to correctly dated evidence. DRAGIN \cite{DRAGIN} triggers retrieval adaptively during generation and formulates queries to reflect the model’s immediate information needs, inserting external knowledge only when beneficial. RAG4DyG \cite{ref:RAG4DyG} retrieves temporally and contextually similar demonstrations from across a dynamic graph and fuses them with the query sequence to adapt to evolving interaction patterns. DynaGRAG \cite{DynaGRAG} enhances subgraph representations through de-duplication, two-step pooling, query-aware retrieval, and dynamic similarity–aware traversal to prioritize diverse yet relevant subgraphs as evidence. GenTKG \cite{ref:GenTKG} retrieves ordered historical events from a temporal knowledge graph as instruction-style context to support generative forecasting of future links. Collectively, these methods show a convergent design: encode temporal structure explicitly, retrieve evidence that is locally time-consistent, and compose temporally ordered contexts so that generation remains faithful to evolving knowledge.
}

\vspace{-3mm}
 {
\section{Practical Applications in Industry} 
Industrial applications of graph-based RAG can be grouped into two tiers: production RAG systems and the workflow–platform stacks. The production tier includes Microsoft's GraphRAG\footnote{https://github.com/microsoft/graphrag}, which builds entity-based knowledge graphs and pre-generated community summaries to improve query-focused summarization; NebulaGraph’s GraphRAG\footnote{https://www.nebula-graph.io/posts/graph-RAG}, which integrates large-language-model reasoning with the NebulaGraph database; Eosphoros’s GraphRAG\footnote{https://github.com/eosphoros-ai/DB-GPT}, which combines DB-GPT \cite{zhou2024db}, OpenSPG\footnote{https://github.com/OpenSPG/openspg}, and TuGraph \footnote{https://tugraph.tech/} for triple extraction, subgraph location, and traversal; Neo4j's NaLLM\footnote{https://github.com/neo4j/NaLLM}, which enables natural-language interfaces to knowledge graphs, construction from unstructured data, and report generation; Writer\footnote{https://writer.com/}, which utilizes graph-based RAG to enhance enterprise content generation and Q\&A systems; and Leroy Merlin\footnote{https://www.lettria.com/case-study/leroy-merlin-knowledge-graph-product-recommendations}, which uses product knowledge graphs to refine recommendation systems, providing personalized customer experiences. 
The workflow–platform tier comprises LangChain\footnote{https://www.langchain.com/} and LangGraph\footnote{https://langchain-ai.github.io/langgraph} for graph-structured pipelines; 
LlamaIndex's Agentic Document Workflows\footnote{https://www.llamaindex.ai/blog/agentic-rag-with-llamaindex-2721b8a49ff6} for end-to-end document processing and meta-agent coordination; 
multi-agent frameworks such as CrewAI\footnote{https://github.com/crewAIInc/crewAI} and AG2 AutoGen\footnote{https://github.com/ag2ai/ag2}; and cloud platforms including Vertex AI\footnote{https://cloud.google.com/vertex-ai}, Amazon Bedrock\footnote{https://aws.amazon.com/cn/bedrock/}, and IBM watsonx\footnote{https://www.ibm.com/products/watsonx}.
}

{\textbf{The Usage of Graph-based RAG in Frontier LLMs.}
Frontier LLMs such as GPT-5.2 \footnote{{https://openai.com/index/introducing-gpt-5-2/}} and Gemini 3 Pro \footnote{{https://ai.google.dev/gemini-api/docs/gemini-3}}  make graph-based RAG more practical because they are stronger at long-context understanding, tool selection, and multi-step planning. Therefore they can reliably act as controllers and verifiers in retrieval pipelines.
In particular, both ecosystems now provide effective retrieval tools named as File Search, which ingest enterprise documents, build searchable indexes, and retrieve only the most relevant evidence for generation \footnote{{https://platform.openai.com/docs/guides/migrate-to-responses}}\footnote{{https://ai.google.dev/gemini-api/docs/file-search}}.
Under our broad definition, this enables graph-based RAG beyond classic knowledge-graph traversal, where document-level structures and hierarchical graphs can be distilled from text and metadata, and retrieval can return compact subgraphs that enforce entity constraints before generation while reducing token and latency costs.
Moreover, Gemini offers explicit controls for extended reasoning, enabling loop and tree style pipelines that iteratively retrieve evidence, verify intermediate results, and revise outputs, which matches the needs of graph-structured pipeline execution in practice \footnote{{https://ai.google.dev/gemini-api/docs/thinking}}. {Compared with earlier LLMs, frontier models mainly address two practical bottlenecks in RAG: they reduce \emph{controller brittleness} for iterative retrieve--verify--revise pipelines, and they make RAG \emph{scalable} to large private corpora by relying on managed retrieval interfaces instead of manual prompt stuffing \cite{li2025long}.}
}

 {
\textbf{Efficiency Discussion.} The efficiency of industrial graph-based RAG can be examined through three views. First, graph construction and precomputation determine the baseline latency and storage footprint: Microsoft's GraphRAG constructs entity-centric knowledge graphs and pre-generates community summaries to speed query-focused summarization at the expense of build time and storage, while Neo4j's NaLLM automates the extraction of nodes, relations, and properties from unstructured data to reduce manual engineering effort in the indexing stage. Second, the retrieval strategy will control most online computation: AntGroup’s GraphRAG narrows work by locating keywords, mapping them to graph nodes, and traversing only the necessary subgraph with breadth-first or depth-first search, and NebulaGraph’s GraphRAG integrates large language models with the graph database to raise precision and avoid wasteful candidate expansion. Third, pipeline design governs coordination overhead and response quality. Sequential chains centralize control and keep latency low, exemplified by LangChain’s modular RAG chains and CrewAI’s support for sequential processes. Loop pipelines add iterative refinement and self-correction when increasing the cost. For example, LangGraph provides explicit loop capability with state persistence so a retriever and generator can revisit intermediate results when needed.
Tree pipelines branch sub-queries or specialized agents for parallel search, which raises coordination cost but can improve answer quality. For instance, AG2 AutoGen together and OpenAI Swarm framework\footnote{https://github.com/openai/swarm} fans out sub-queries before synthesis, which improves coverage at the cost of additional coordination.
}

\section{Future Research Opportunities}

\textbf{Adaptive Prompts for LLMs.} Future research could delve into the design and evaluation of adaptive prompts to better align with the pretrained knowledge of LLMs. By tailoring the structure, terminology, and hierarchy of information drawn from graphs, system designers could harness improved synergy between graph-based data and the LLM’s reasoning capabilities. One could also investigate methods to prioritize the most critical entities, relationships, or paths in a query~\cite{ref:gnn-ret, ref:kg-rank}, ensuring that the prompt remains precise and contextually rich without overwhelming the model. Such selective attention could enhance both the relevance and interpretability of LLM outputs. Furthermore, exploring feedback mechanisms that dynamically refine the prompt based on the LLM’s responses might be an interesting direction, allowing systems to continuously optimize for coherence and clarity. Ultimately, adaptive prompt strategies could pave the way for more accurate, context-aware question answering, harnessing the potential of complex graph structures in tandem with cutting-edge language models.

\textbf{Enhanced Understanding for Graph Problems.} Future research could delve into approaches that enable deeper structural comprehension of graphs within LLMs. Since many current models rely on sequential token representations, they struggle to handle larger or more complex graph structures—especially when a sequence involves a large number of nodes~\cite{graphwiz, ref:GPT4Graph}. This limitation hinders their ability to solve typical graph-centric problems, such as path finding or community detection, which often require more specialized, non-linear reasoning. One direction might be to integrate graph-specific modules or adopt novel representations that faithfully capture relational information without overwhelming the model. Moreover, another avenue of exploration could involve combining LLMs with algorithmic components adopted at handling extensive, node-rich graphs.

\textbf{Multi-Modal Graphs for Cross-Domain RAG.} 
Future advancements in Multi-Modal Graphs for Cross-Domain RAG should focus on the development of adaptable retrieval algorithms that accommodate the diversity of data types encountered in real-world applications. Presently, graph-based RAG systems are largely confined to single-domain use cases, often employing retrieval methods tailored to specific types of textual information. However, as cross-domain scenarios become increasingly common, there is a pressing need for retrieval techniques that can seamlessly handle multi-modal graph structures, including images \cite{han2022vision}, audio \cite{nie2020c}, and numerical data. These enhanced retrieval algorithms must not only integrate and interpret a wide range of data formats but also dynamically adjust their retrieval strategies to match the unique characteristics of each domain. By enabling a more flexible, multi-modal approach, future research in this field can overcome the limitations of current systems, paving the way for more robust and versatile knowledge acquisition across diverse domains.


\textbf{Improved Graph Construction Techniques.} The future of RAG systems can center around overcoming the limitations of traditional triple-based knowledge graphs, which are often too simplistic for complex, real-world data ~\cite{ref:noname8}. While the subject-predicate-object \cite{zhang2017visual} format has served as a foundation, it does not fully capture intricate relationships and evolving patterns within knowledge. To address this, future graph construction methods should explore hypergraphs, which allow for multi-node relationships \cite{li2007network}, and semantic embeddings \cite{wang2018zero} that can represent context, enhancing the depth and accuracy of graph structures. Additionally, GNNs can be used to learn graph representations that dynamically evolve as new information is introduced, allowing for more precise and adaptive knowledge storage. Another important direction will be the development of hierarchical graphs such as the approaches utilized in \cite{ying2018hierarchical} that capture different levels of abstraction, providing a more flexible and structured way to organize knowledge. Furthermore, adaptive graph construction techniques, which adjust the structure and organization of the graph based on the type and volume of data, will be essential for handling real-time updates without sacrificing retrieval speed. In summary, these advanced techniques will allow RAG systems to store knowledge more efficiently, retrieve information faster, and support more complex, multi-modal data interactions.


\textbf{Incorporating User Interaction.} Future work in RAG systems can be significantly improved by integrating user interaction to improve the efficiency, accuracy, and user satisfaction of knowledge retrieval. By integrating Human-Computer Interaction (HCI) principles, RAG systems can offer more intuitive and adaptive interfaces, allowing users to actively shape the retrieval process. For example, adaptive query refinement \cite{deshpande2007adaptive} could enable users to iteratively clarify or expand their queries, with the system responding to these modifications in real-time. Additionally, interactive visualizations of the knowledge graph could allow users to explore relationships between data points, enhancing understanding and control over the retrieval process. Incorporating user feedback loops \cite{mansoury2020feedback} — where users can rate the relevance of results or suggest corrections—would also help the system learn and adapt over time, making it more responsive to evolving user needs. Moreover, adopting context-aware systems \cite{hong2009context} could allow RAG systems to interpret not only the user's query but also the broader context of the interaction, further refining results based on user preferences and past behaviors. By blending these HCI concepts into RAG design, future systems can offer more personalized, efficient, and transparent experiences, ultimately enhancing user trust and satisfaction.




\section{Conclusion}
This paper provides a comprehensive survey of the integration of graph-based techniques into RAG systems, offering a detailed review of their applications, advancements, and challenges. By categorizing existing methods and proposing a novel taxonomy, it provides valuable insights into how graphs can enhance the reasoning capabilities and accuracy of LLMs. The survey also identifies key challenges, such as dynamic graph integration, scalability, and explainability, while outlining future research directions to address these issues. Overall, this work contributes to a deeper understanding of the role of graphs in RAG systems and lays a foundation for future research aimed at unlocking their potential in improving LLM performance across a wide range of tasks.












\bibliographystyle{acm}  
\bibliography{main}   
\clearpage
\newpage

\section{Appendix}

\subsection{Brief Introduction of Works in Table \ref{table:task}}
\textbf{ReTraCk}~\cite{ref:retrack} utilizes a retriever, a transducer, and a checker to parse questions into syntax guaranteed logical forms. It retrieves entities and schema, applies a grammar based decoder, and performs instance level and ontology level checks with execution, thus reaching top accuracy with interactive latency on GrailQA and WebQuestionsSP.

\textbf{Golden-Retriever}~\cite{ref:golden} utilizes reflection based question augmentation before retrieval by identifying jargon, inferring context, and rewriting the query. It also enriches the document store through optical character recognition and language model summarization, thus improving retrieval fidelity in industrial knowledge bases.

\textbf{Fact}~\cite{ref:fact} utilizes efficient subgraph retrieval for fact verification over knowledge graphs and compares textual fine tuning, a question answering graph neural network hybrid, and large language model prompting. It reports higher accuracy and faster training on the FACTKG dataset. 

\textbf{RNG-KBQA}~\cite{ref:rng} utilizes a rank then generate pipeline in which a BERT based bi encoder scores logical form candidates and a T5 based sequence to sequence generator composes the final logical form. It achieves new state of the art results on GrailQA and WebQSP.

\textbf{TOG2}~\cite{ref:tog2} alternates between graph retrieval and context retrieval to discover in-depth clues, and it tightly couples knowledge graphs with text so that each side improves the other. It aims for deep and faithful reasoning, and it remains training free and plug and play while delivering state-of-the-art accuracy on knowledge-intensive tasks.

\textbf{MINERVA}~\cite{ref:minerva} learns a policy to walk on a knowledge graph conditioned on a query, and it answers by reaching the correct node. It supports variable-length paths, yields interpretable reasoning traces, and improves efficiency by searching local neighborhoods rather than ranking all entities.

\textbf{KALMV}~\cite{ref:kalmv} introduces a small instruction-tuned verifier that checks whether retrieved knowledge is relevant and whether the generated answer is grounded, and it rectifies errors by re-retrieving or regenerating when needed. It works as a plug and play module and reduces hallucinations across open-domain and knowledge-graph question answering.

\textbf{LLaGA}~\cite{ref:llaga} adapts graphs to large language models by reorganizing nodes into structure-aware sequences and mapping them into the token embedding space through a projector. It preserves the general-purpose nature of language models and shows strong performance and interpretability across datasets and tasks.

\textbf{ROG}~\cite{ref:rog} proposes a planning, retrieval, and reasoning framework. It first generates relation paths as faithful plans grounded in the knowledge graph, then retrieves reasoning paths so the model answers with interpretable explanations.

\textbf{RRA}~\cite{ref:rra} introduces a retrieve, rewrite, and answer pipeline. It rewrites retrieved subgraphs into answer sensitive knowledge text with a KG to Text module and uses the text to guide the question answering model, which improves KGQA.

\textbf{KnowledgeNavigator}~\cite{ref:kn} builds a three stage KGQA process. It predicts hop count and generates similar questions to constrain search, then iteratively retrieves a focused subgraph and composes a reasoning prompt to produce the answer.

\textbf{KD CoT}~\cite{ref:kd-cot} turns Chain of Thought into a structured multi round dialogue with an external QA system. It verifies and amends intermediate steps using retrieved answers and thus yields more faithful reasoning and stronger results on WebQSP and ComplexWebQuestions.

\textbf{MVP Tuning}~\cite{ref:mvp-tuning} combines multi view knowledge retrieval with parameter efficient prompt tuning. It augments the input with self view and consensus view triplets and jointly models text and knowledge in a single prompt tuned language model, which delivers state of the art accuracy on several commonsense QA benchmarks.

\textbf{KagNet}~\cite{ref:kagnet} constructs a schema graph from an external commonsense graph and models it with a knowledge aware graph network that combines graph convolutional networks and LSTMs with hierarchical path attention, and it yields transparent reasoning and state of the art results on CommonsenseQA.

\textbf{KG-Agent}~\cite{ref:kg-agent} builds an autonomous agent that integrates a language model with a toolbox, a knowledge graph executor, and a memory, and it iteratively selects tools and updates memory to complete multi hop reasoning while a small tuned model surpasses stronger baselines on in domain and out of domain benchmarks.

\textbf{GenKGQA}~\cite{ref:gentkgqa} proposes a two stage generative framework for temporal knowledge graph question answering that first retrieves a question relevant temporal subgraph using language model guidance and then generates answers with instruction tuned fusion of graph signals and text through virtual knowledge indicators.

\textbf{UniKGQA}~\cite{ref:unikgqa} unifies retrieval and reasoning with a semantic matching module and a matching information propagation module, and it introduces abstract subgraphs and shared pre training and fine tuning so that it achieves large improvements on WebQSP and ComplexWebQuestions.

\textbf{Knowledge Solver}~\cite{ref:ksl} teaches language models to search a knowledge graph by turning retrieval into a multi hop decision sequence and by producing complete reasoning paths, and it improves zero shot and fine tuned performance on CommonsenseQA, OpenbookQA, and MedQA while increasing explainability.

\textbf{WebQSP}~\cite{ref:webqsp} provides SPARQL labeled semantic parses for questions from WebQuestions and shows that such labels can be collected efficiently, and it demonstrates that training with semantic parses yields about five points higher average F1 and more consistent and updatable answers.

\textbf{MetaQA}~\cite{ref:metaqa} proposes a variational reasoning network that jointly handles uncertain topic entities and multi hop reasoning through a probabilistic framework, and it introduces a propagation style architecture trained with variance reduced reinforcement learning. It also releases a large scale benchmark of more than four hundred thousand questions across one hop to three hop settings and shows strong results on text and audio queries.

\textbf{PullNet}~\cite{ref:pullnet} performs iterative retrieval to build a question specific subgraph and then reasons over this compact graph with a graph convolutional module. It learns which nodes to expand at each step and achieves large gains on multi hop datasets. 

\textbf{KnowledGPT}~\cite{ref:knowledgpt} integrates large language models with external knowledge bases by generating program of thought search code to call KB functions, then executing the retrieved results for answering. It supports both retrieval and storage over public and private KBs without task specific training, and it differs from classical KBQA systems by covering diverse knowledge types.

\textbf{DecAF}~\cite{ref:decaf} jointly decodes logical forms and direct answers, and then combines the two signals to produce final predictions. It also linearizes knowledge bases into text to use general purpose retrieval, which improves adaptability across datasets.

\textbf{GGE}~\cite{ref:gge} partitions a retrieved knowledge subgraph into smaller subgraphs and applies a graph augmented learning to rank model that captures global and local interactions between the question and subgraphs. It reduces search space and benefits subsequent answer selection in information retrieval based KGQA.

\textbf{NSM}~\cite{ref:nsm} adopts a teacher student design that couples forward and backward reasoning to learn reliable intermediate supervision signals and improve multi hop KBQA. It aligns entity distributions from both directions to mitigate spurious paths and strengthen the student model. 

\textbf{SKP}~\cite{ref:skp} introduces structured knowledge aware pre training to narrow the gap between text and subgraphs. It adds an interval attention mechanism and an efficient linearization strategy, which improve retrieval and robustness under incomplete knowledge bases.

\textbf{KnowGPT}~\cite{ref:knowgpt} builds a knowledge graph based prompting framework for large language models. It extracts informative knowledge and constructs context aware prompts to reduce hallucination and raise accuracy on knowledge intensive tasks.

\textbf{DiFaR}~\cite{ref:difar} performs direct fact retrieval by embedding facts and queries in one space and retrieving triplets without entity linking. It then reranks top candidates, which avoids error propagation and yields strong results while remaining simple.

\textbf{KAPING}~\cite{ref:kaping} augments zero shot prompting by retrieving relevant triples, verbalizing them as text, and injecting them into the prompt so answers are grounded in factual knowledge. It filters to the most similar triples and shows consistent gains across many language models.

\textbf{NuTrea}~\cite{ref:nutrea} adopts a neural tree search scheme for multi hop knowledge graph question answering. It uses expansion, backup, and node ranking with a relation frequency inverse entity frequency embedding to incorporate broader graph context and improve path selection.

\textbf{OREOLM}~\cite{ref:oreolm} inserts knowledge interaction layers into a transformer to interact with a knowledge graph reasoning module through contextualized random walks. It strengthens closed book question answering and provides interpretable reasoning paths.

\textbf{GPT4Graph}~\cite{ref:GPT4Graph} evaluates how large language models understand graph structured data across structural and semantic tasks, and it builds a benchmark while identifying current limits and future directions.

\textbf{GraphText}~\cite{ref:graphtext} introduces a graph syntax tree that converts graphs into natural language prompts so an LLM can reason in text without extra training, and it injects graph inductive biases for stronger reasoning and explanation.

\textbf{ENGINE}~\cite{ref:engine} presents an efficient tuning and inference framework for textual graphs by linking frozen language model layers with graph ladders in a side structure, and it uses caching and dynamic early exit to reduce time and memory while keeping accuracy high.

\textbf{GraphBridge}~\cite{ref:graphbridge} proposes a multi granularity integration that bridges local node text and global topology through contextual textual information, and it adds a graph aware token reduction module for scalable learning with state of the art results.

\textbf{MMGCN}~\cite{ref:mmgcn} constructs a user item bipartite graph for each modality and propagates signals to capture modal specific preferences, and it shows clear gains on Tiktok, Kwai, and Movielens.

\textbf{ConvE}~\cite{ref:wn} applies two dimensional convolutions over embeddings for knowledge graph link prediction with a simple yet expressive architecture, and it reaches state of the art mean reciprocal rank with efficient one to many scoring.

\textbf{Graph-LLM}~\cite{ref:Graph-LLM} explores two pipelines for text-attributed graphs, using large language models either to enhance node attributes alongside graph neural networks or to act as direct predictors for node classification; it finds enhancer pipelines effective while predictor pipelines show preliminary promise yet risk inaccurate outputs and evaluation leakage.

\textbf{NLGraph}~\cite{ref:nlgraph} introduces a natural-language benchmark of 29,370 problems across eight graph reasoning tasks and tests whether prompting strategies improve model reasoning; it shows gains on simple tasks with chain-of-thought style prompting, yet the benefits diminish and can reverse on complex problems.

\textbf{GLBench}~\cite{ref:GLBench} provides a comprehensive benchmark for GraphLLM methods in supervised and zero shot settings, unifying datasets and splits to compare enhancer, predictor, and aligner roles; it reports that enhancer methods are robust, predictor methods are less reliable, and a training-free baseline can be competitive. 

\textbf{GraphEval2000}~\cite{ref:GraphEval2000} constructs a code-based evaluation suite with 40 graph problems and 2,000 test cases and adds structured symbolic decomposition to guide solutions; it returns failed cases for feedback and shows sizable gains for popular models under the proposed instruction design.

\textbf{MuseGraph}~\cite{ref:musegraph} proposes a graph-oriented instruction-tuning framework that builds compact graph descriptions, generates diverse chain-of-thought instructions, and applies graph-aware tuning; it achieves superior results across five tasks and ten datasets and supports dynamic allocation of instruction volume by task and dataset complexity.

\textbf{KG-Rank}~\cite{ref:kg-rank} integrates a medical knowledge graph with ranking and re ranking to support long form medical question answering. It identifies entities, retrieves related triples, and orders them to improve factuality and relevance for the final answer.

\textbf{FABULA}~\cite{ref:fabula} constructs an event plot graph from open source reports and guides stepwise retrieval and reasoning for intelligence analysis. It then drafts structured situation reports with evidence tracing and explanation.

\textbf{FoodGPT}~\cite{ref:foodgpt} is a domain large language model for food testing that adds incremental pre training and instruction tuning. It integrates an external knowledge graph for retrieval to reduce hallucination and to support precise outputs. 

\textbf{KAM CoT}~\cite{ref:kamcot} trains a light verifier to check retrieval relevance and grounding during knowledge augmented generation. It detects retrieval errors and grounding errors and it triggers corrective actions to improve reliability. 

\textbf{MufassirQAS}~\cite{alan2024rag} curates Quranic sources and builds a retrieval augmented pipeline with knowledge driven summarization. It ranks and synthesizes evidence to produce faithful answers with clarity for non expert users.

\textbf{MEDQA}~\cite{ref:medqa} is an open domain multiple choice dataset drawn from medical board examinations in the United States, Mainland China, and Taiwan. It requires retrieval over medical textbooks and logical reasoning for professional decisions.

\textbf{DepsRAG}~\cite{ref:depsrag} is a multi agent RAG system for software dependency analysis that builds a dependency graph and answers queries over it. It coordinates assistant, search, dependency, and critic agents to decompose tasks and to validate outputs.

\subsection{Additional Related Works}

\textbf{GLBK} \cite{ref:noname5}  introduces a novel retrieval method for biomedical knowledge that leverages a knowledge graph to address the limitations of embedding-based similarity retrieval. This method constructs a biomedical knowledge graph by utilizing entity recognition and relationship extraction techniques, specifically fine-tuning PubmedBERT \cite{ref:pubmedbert} for the biomedical domain. This graph builds text chunks from the PubMed \footnote{https://ftp.ncbi.nlm.nih.gov/pubmed/} corpus by associating them with nodes (entities like genes, diseases, or compounds) and edges (relationships). Unlike embedding-based retrieval methods, which tend to prioritize over-represented clusters in the data, this graph-based approach re-balances information by under-sampling dense clusters and ensuring access to long-tail, less frequent yet valuable knowledge. By combining the semantic strengths of embedding similarity with the structural insights from the knowledge graph, the proposed hybrid model demonstrates superior performance of accuracy, addressing the information overload problem prevalent in biomedical research retrieval.

\textbf{ATLANTIC} \cite{ref:atlantic} is a structure-aware retrieval-augmented language model that addresses limitations in current RAG systems by integrating document structural relationships. A heterogeneous document graph is constructed from scientific literature, capturing four types of relationships: co-citation, co-topic, co-venue, and co-institution, to connect documents across 15 scientific disciplines. This structure is used to encode the relationships with a pre-trained Heterogeneous Graph Transformer (HGT) \cite{hu2020heterogeneous}, which generates structural embeddings. These embeddings are fused with semantic embeddings from text, enabling ATLANTIC to retrieve passages that are both semantically relevant and structurally coherent. The approach mitigates the shortcomings of text-only retrieval methods, particularly in interdisciplinary domains where relational context is crucial. The model demonstrates improved faithfulness and relevance in retrieving passages for scientific tasks like question answering and document classification.

\textbf{GNN-Ret} \cite{ref:gnn-ret} constructs a database tailored for improved passage retrieval in complex question-answering scenarios. The authors create a "Graph of Passages" (GoPs) by linking textual passages based on two key principles: structural relationships and shared keywords. Structural connections are established by linking adjacent passages in documents, maintaining their order and context. Keyword connections are derived by extracting entities using LLMs, and then linking passages that share the same keywords to highlight semantic associations. This graph-based representation addresses challenges like retrieving semantically distant but related passages, enabling improved retrieval coverage and more accurate question answering. The innovative use of GoPs mitigates the limitation of traditional database, which often treat passages as isolated units, and introduces an efficient way to model and retrieve interconnected information.


\textbf{Explore-then-Determine (EtD)} \cite{ref:etd}  utilizes a graph convolutional method to enhance graph retrieval by employing adaptive propagation and message passing. In the \textit{Explore} phase, the framework initializes with a topic entity and uses a lightweight GNN, integrated with semantic representations from LLMs, to expand the candidate set by propagating to relevant neighbors. Attention weights are computed to rank and prune irrelevant edges, focusing on the top-$K$ edges at each layer to refine candidate sets. In the \textit{Determine} phase, the filtered graph is processed by a frozen LLM with a knowledge-enhanced multiple-choice prompt, merging explicit graph-based knowledge with the LLM’s implicit understanding to determine the final answer. This approach efficiently filters irrelevant KG information while enabling precise multi-hop reasoning.


\textbf{GRAG \cite{ref:grag}} introduces a dual approach to graph-structure prompting for integrating textual graphs into LLMs. In the Graph View, textual graphs are encoded as soft prompts using a Graph Neural Network (GNN) to preserve topological information. Retrieved subgraph embeddings are aligned with LLM token embeddings via an MLP, ensuring seamless integration of graph structure. In the Text View, textual graphs are transformed into hierarchical text descriptions as hard prompts. This involves splitting retrieved ego-graphs into a \textit{Breadth-First Search (BFS) tree} and residual edge set, then organizing them into structured templates to retain narrative and topological context. The query, hierarchical text description, and graph embeddings are concatenated as input to the LLM, enabling context-aware generation that leverages both structural and semantic aspects of the graph.

\textbf{MindMap \cite{ref:mindmap}} creates prompts for LLMs by converting knowledge graph subgraphs into natural language descriptions. First, it mines evidence subgraphs using path-based and neighbor-based exploration techniques to gather relevant information. Then, it performs an evidence graph aggregation step, where each subgraph is formatted as an entity chain, such as $\mathcal{G}' = \{(e_h, r, e_t)\}$, which is translated into a natural sentence by a predefined template. These descriptions are merged to form a reasoning graph, which is included as part of the LLM's prompt. This approach ensures that the LLM understands both the structural and semantic aspects of the graph, enabling robust and context-aware reasoning.

\textbf{KGP \cite{ref:kgp}} formulates the process of generating knowledge graph prompts for multi-document question answering scenarios into two steps. First, a general knowledge graph is constructed based on the passage similarity from multiple documents. During the process, the merits and limitations of different construction methods are examined, to build the mapping between the most appropriate methods and scenarios. Second, a language model agent is employed to traverse the knowledge graph and retrieve relevant context. The agent traverses over the KG through the new evidence given by the LLM iteratively according to the retrieved passages. The final structured passage queue from the KG services acts as the prompt for multi-document question answering.


\textbf{ArcaneQA} \cite{ref:arcaneqa} is designed to address the combinatorial explosion problem commonly encountered in searching for matched subgraphs within large-scale knowledge bases. Unlike traditional approaches that involve an exhaustive search, ArcaneQA narrows its focus to a small set of admissible tokens derived from the relevant subgraph, rather than considering the entire vocabulary. This selective prediction significantly reduces computational complexity. Utilizing an encode-decode architecture built on top of the BERT model, ArcaneQA iteratively predicts additional sequences from the subgraph, allowing for efficient stepwise progression. By breaking down the task of program generation into a sequence of smaller, more manageable decision-making processes, ArcaneQA effectively transforms a daunting search challenge into a tractable series of localized predictions, streamlining the reasoning process while maintaining accuracy.

\textbf{EWEK-QA} \cite{ref:ewek-qa} triggers two pipelines simultaneously to extract informative knowledge from the web content and the knowledge base. For the retrieval from the web text,  EWEK-QA first retrieves the relevant content utilizing the Bing search and then conducts a series of processes to filter and rerank the candidate texts from the web pages. Simultaneously, EWEK-QA avoids retrieval techniques relying on LLMs such as TOG \cite{ref:tog} and directly employs a match of representation similarity (cosine score) to obtain the relevant triples. To this end, EWEK-QA further utilizes a pre-trained LLMs to integrate KG triples and web quotes together and construct a unified prompt. By substituting the iterative retrieval in the knowledge base and enhancing it with web search, EWEK-QA significantly improves the quality of extracted knowledge in terms of relevance to the queries without hampering efficiency.

\textbf{SR \cite{ref:sr}} aims to enhance  KGQA by efficiently retrieving high-quality subgraphs closely aligned with the query. It achieves this by decoupling the retrieval process from the reasoning step, enabling a plug-and-play framework that can be integrated with various KGQA models. SR uses a dual-encoder architecture to expand and refine paths, automatically stopping when relevant subgraphs are constructed. This separation reduces reasoning bias caused by partial subgraphs and improves retrieval precision. By employing weakly supervised and end-to-end fine-tuning strategies, SR demonstrates significant improvements in reasoning accuracy, explainability, and retrieval efficiency, achieving state-of-the-art results in complex multi-hop KGQA tasks.

\textbf{LLMs-EP} \cite{ref:Graph-LLM} focuses on understanding how text-attribute GNNs can leverage LLMs to enhance node classification tasks. This paper presents an empirical study of two distinct pipelines for incorporating LLMs into graph tasks. In the first pipeline, LLMs are employed as enhancers targeting the text content associated with each node. Specifically, LLMs-EP utilizes LLMs either as embedding encoders to generate node representations or to incorporate additional textual content from related nodes. These approaches are termed feature-level and text-level enhancement, respectively, enabling more informative representations of nodes. In the second pipeline, LLMs-EP provides an in-depth analysis of two scenarios: first, exploring whether LLMs can directly predict node categories, and second, examining the impact of using 2-hop neighbors to incorporate structural information into the prediction process. By combining text and structural data, LLMs-EP aims to improve the accuracy and understanding of node classification in graph-based tasks.

\end{document}